\documentclass[aps,10pt,twocolumn,showpacs,pra,amsmath,amssymb,floatfix,superscriptaddress,dvipsnames]{revtex4-2}

\usepackage{amsthm}
\usepackage{mathtools}
\usepackage{physics}
\usepackage{bbm}
\usepackage{amsfonts}
\usepackage{amssymb}
\usepackage{graphicx}
\usepackage{xfrac}
\usepackage[T1]{fontenc}
\usepackage{bbold}
\usepackage{float}
\usepackage[utf8]{inputenc}
\usepackage{csquotes}
\usepackage{xcolor}
\definecolor{color_scheme}{RGB}{26, 121, 199}
\usepackage[colorlinks=true,linktocpage=true]{hyperref}
\hypersetup{
    allcolors  = {color_scheme},
}

\theoremstyle{plain}

\usepackage{array}
\usepackage{multirow}

\usepackage[utf8]{inputenc}
\usepackage[english]{babel}
\usepackage[T1]{fontenc}

\usepackage{amsfonts}
\usepackage{dsfont}

\usepackage{enumerate}

\usepackage{braket}
\usepackage{framed}
\usepackage{graphicx}

\usepackage{hyperref}
\usepackage{cleveref}
\usepackage{natbib}
\usepackage{csquotes}
\usepackage{xcolor}
\usepackage[normalem]{ulem}

\usepackage{bbold}

\newcommand{\inlshort}{INL -- International Iberian Nanotechnology Laboratory, Braga, Portugal} 

\newcommand{\pisashort}{Department of Physics ``E. Fermi'', University of Pisa, Pisa, Italy} 

\newcommand{\cfumshort}{Centro de F\'{i}sica, Universidade do Minho, Braga, Portugal} 

\newcommand{\uffshort}{Instituto de Física, Universidade Federal Fluminense, Niterói—RJ, Brazil} 

\newcommand{\uwienshort}{Fakultät für Mathematik, Universität Wien, Vienna, Austria} 

\newcommand{\ulmshort}{Institut für Theoretische Physik, Universität Ulm, Ulm, Germany} 

\usepackage{soul}

\begin{document}

\title{Estimation of multivariate traces of states given partial classical information}

\author{Kyrylo Simonov}
\email{kyrylo.simonov@univie.ac.at} 
\affiliation{\uwienshort}

\author{Rafael Wagner}
\email{rafael.wagner@uni-ulm.de} 
\affiliation{\inlshort}
\affiliation{\cfumshort}
\affiliation{\pisashort}
\affiliation{\ulmshort}

\author{Ernesto Galvão}
\email{ernesto.galvao@inl.int} 
\affiliation{\inlshort}
\affiliation{\uffshort}

\date{\today}

\begin{abstract}
Bargmann invariants of order $n$, defined as multivariate traces of quantum states $\text{Tr}[\rho_1\rho_2 \ldots \rho_n]$, are useful in applications ranging from quantum metrology to certification of nonclassicality. A standard quantum circuit used to estimate Bargmann invariants is the cycle test. In this work, we propose generalizations of the cycle test applicable to a situation where $n$ systems are given and unknown, and classical information on $m$ systems ($m\leq n)$ is available, allowing estimation of invariants of order $n+m$. Our main result is a generalization of results on 4th order invariants appearing in `double' weak values from~\href{https://journals.aps.org/prresearch/abstract/10.1103/PhysRevResearch.6.043043}{Chiribella~et al. [Phys. Rev. Research 6, 043043 (2024)]}. The  use of classical information on some of the states enables circuits on fewer qubits and with fewer gates, decreasing the experimental requirements for
their estimation, and enabling multiple applications we briefly review.
\end{abstract}

\maketitle

\section{Introduction}

Bargmann invariants~\cite{bargmann1964note, simon1993Bargmann, mukunda2001Bargmann, mukunda2003Bargmann, mukunda2003Wigner, akhilesh2020geometric} are multivariate traces of quantum states. These are defined as
\begin{equation}\label{eq: Bargmann invariant}
    \Delta_{n}(\pmb\varrho) = \text{Tr}[\rho_1\rho_2 \ldots \rho_n],
\end{equation}
where $n$ is defined to be the \emph{order} of the invariant and $\pmb\varrho \equiv (\rho_1,\ldots,\rho_n) \in \mathcal{D}(\mathcal{H})^n$ is a tuple of quantum states, where $\mathcal{D}(\mathcal{H})$ denotes the set of density matrices associated to the Hilbert space $\mathcal{H}$. The values $\Delta_n(\pmb\varrho)$ are invariant under the transformation 
\begin{equation}\label{eq: unitary transformation}
(\rho_1,\rho_2,\ldots,\rho_n) \mapsto (U \rho_1 U^\dagger, U \rho_2 U^\dagger, \ldots, U \rho_n U^\dagger)
\end{equation}
for any unitary $U: \mathcal{H}\to \mathcal{H}$. 

The investigation of generic multivariate traces of matrices for understanding quantum mechanical properties of physical systems goes back to the very beginning of quantum theory,  e.g. in $n$-point correlation functions for quantum statistical mechanics~\cite{fano1957description}. These multivariate traces of quantum states have multiple applications, motivating the development of efficient ways for measuring them. In photonics, their phases~\footnote{If we write $\Delta_n(\pmb\varrho) = \Delta e^{i\phi_\Delta}$, with $\Delta \in \mathbb{R}_{\geq 0}$, the value $\phi_\Delta$ is called the \emph{phase} of the invariant.} are known as collective photonic phases~\cite{shchesnovich2015partial,shchesnovich2018collective}, and have been measured for invariants of up to fourth order~\cite{bong2018strong,jones2020multiparticle,pont2022quantifying,jones2023distinguishability}. They are essential for the complete characterization of multiphoton indistinguishability~\cite{rodari2024experimentalobservationcounterintuitivefeatures,rodari2024semideviceindependentcharacterizationmultiphoton}, as required by linear-optical photonic quantum computation~\cite{shchesnovich2015partial,oszmaniec2024measuring}.

Already the simplest Bargmann invariants have their utility. Second order invariants $\text{Tr}[\rho_1\rho_2]$, also known as two-state overlaps, have been shown to provide sufficient information for witnessing various nonclassical resources of quantum states such as quantum coherence~\cite{galvao2020quantumandclassical,giordani2021witnessing,wagner2024inequalities,wagner2024coherence}, nonstabilizerness~\cite{wagner2024certifying,haug2023scalable}, preparation contextuality~\cite{wagner2024coherence,giordani2023experimentalcertification}, Kochen--Specker contextuality~\cite{wagner2024inequalities}, indistinguishability~\cite{brod2019witnessing,giordani2020experimentalquantification,giordani2021witnessing}, and Hilbert space dimension~\cite{giordani2021witnessing,giordani2023experimentalcertification}. Two-state overlaps can be semi-device independently self-tested~\cite{miklin2021universalscheme} and be used for certifying properties of distributed quantum computers~\cite{hinsche2024efficientdistributedinnerproduct}, symmetry of quantum channels~\cite{bandyopadhyay2023efficient}, and for performing efficient fidelity-based certification of almost all quantum states~\cite{huang2024certifyingquantumstatessinglequbit}. 

There are various applications requiring Bargmann invariants of order $n > 2$. Third-order invariants describe weak values~\cite{tamir2013introduction,dressel2014colloquium,wagner2023simple} and standard Kirkwood--Dirac (KD) quasiprobability representations~\cite{lostaglio2023kirkwooddirac,wagner2024quantum,gherardini2024quasiprobabilities,arvidssonshukur2024properties,schmid2024kirkwooddirac,liu2025boundarykirkwooddiracquasiprobability}. Some experimentally relevant situations actually require the estimation of invariants of order $n>3$. For example, consider four pure states $\{\vert \phi_1\rangle \langle \phi_i|\}_{i=1}^4$ such that $\langle \phi_1|\phi_3\rangle = \langle \phi_2 \vert \phi_4\rangle = 0$. In such a case, all third-order invariants $\Delta_{3}(\phi_i,\phi_j,\phi_k)$ are equal to zero yet the fourth-order invariant may be non-null. This implies that this invariant is \emph{not} a function of the other third-order invariants. Moreover, as pointed out in Ref.~\cite{oszmaniec2024measuring}, in general when states are not pure one {must} estimate multivariate traces of states of large orders in order to completely characterize the relational information classes (i.e., completely solve the unitary invariance problem). Other situations where one \emph{must} estimate higher-order invariants of (in general) mixed quantum states are for spectrum estimation $\{\text{Tr}[\rho^k]\}_{k}$~\cite{alves2003direct,brun2004measuring,leifer2004measuring,vanenk2012measuring,tanaka2014determining,wagner2024quantum,shin2024rankneedestimatingtrace}, {entanglement spectroscopy}~\cite{johri2017entanglement}, quantifying entanglement and nonstabilizerness~\cite{turkeshi2023measuring,tirrito2024quantifying}, characterizing topological order, quantum phase transitions and emergent irreversibility~\cite{dechiara2012entanglement,chamon2014emergent}, performing quantum error mitigation~\cite{koczor2021exponential,huggins2021virtual}, Gibbs state preparation~\cite{wang2021variational}, and virtual cooling~\cite{cotler2019cooling}. Measurements of 5th-order invariants $\text{Tr}[\rho_1\rho_2\rho_3\rho_4\rho_5]$ enable the characterization of the degree of incompatibility between observables~\cite{gao2023measuring}, and of information scrambling and quantum chaos via out-of-time-ordered correlators (OTOCs)~\cite{halpern2018quasiprobability,wagner2024quantum,gonzales2019otoc,gonzales2022diagnosing}. Invariants of even higher order appear in the description of sequential weak measurements~\cite{mitchison2007sequential} and of the Pancharatnam phase~\cite{pancharatnam1956generalized,arvind1997generalized} acquired after a sequence of projective measurements~\cite{wilczek1989geometric,chruciski2004geometric,fernandes2024unitaryinvariant}. It is well known that the Pancharatnam phase is connected to geometric phases, also known as Berry phases~\cite{berry1984quantal,avdoshkin2023extrinsic,simon1993Bargmann,akhilesh2020geometric,mukunda2001Bargmann,mukunda2003Bargmann,mukunda2003Wigner,rabei1999Bargmann}. 

Having clarified the many applications provided by estimating $n$-th order Bargmann invariants for which $n \geq 3$, we note that state of the art quantum circuits aimed at measuring $\Delta_n(\pmb\varrho)$ require the application of at least $n$ Fredkin gates on at least $n$ qubits~\cite{quek2024multivariatetrace,oszmaniec2024measuring}. This protocol for estimating Bargmann invariants is known as the \emph{cycle test} protocol. A detailed comparison between estimating $\Delta_n(\pmb\varrho)$ using the cycle test and other approaches can be found in Ref.~\cite{wagner2024quantum}. The current implementations of the cycle test quantum circuits assume \emph{no} prior information of the tuple $\pmb\varrho$. The states in $\pmb\varrho$ can be unknown inputs, and the estimation works even when no prior classical description---e.g. their complete tomographic information, or a classical description of a state preparation circuit---is available. In other words, the quantum circuits  from Refs.~\cite{quek2024multivariatetrace,oszmaniec2024measuring}  work even if the input states are provided by another party. This raises the question: How can prior classical information improve the estimation of invariants using the cycle test?

Recently, Chiribella et al.~\cite{chiribella2024dimensionindependentweakvalueestimation} showed, using a variation of the swap test, that it is possible to estimate the quantity $\text{Tr}[\rho_1 A \rho_2 B]$, where $A$ and $B$ are two known observables. This implies that choosing $A$ and $B$ to be related to quantum states the same circuits can estimate fourth order invariants. This significantly simplifies tests of such higher-order Bargmann invariants provided that one has the ability to perform some degree of classical postprocessing on the observed data, and provided classical information on a subset of the quantum states is available. 

\begin{figure}[t]
    \centering
    \includegraphics[width=\columnwidth]{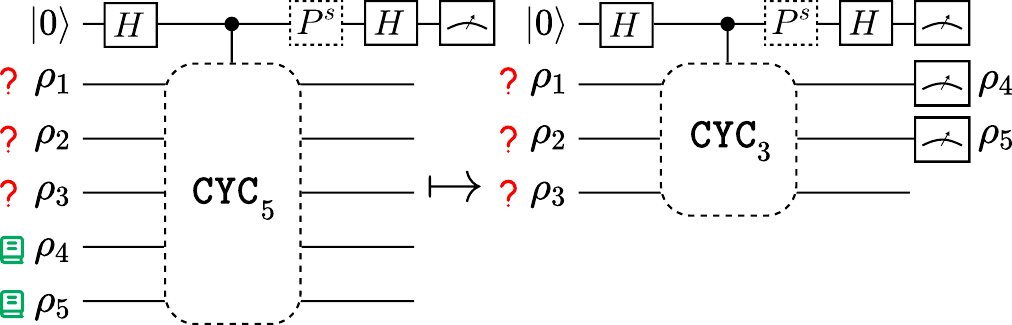}
    \caption{\textbf{Main result.} Quantum circuits for estimating multivariate traces---expressed as $\Delta_n(\pmb\varrho) = \text{Tr}[\rho_1 \ldots \rho_n] = \text{Tr}[\mathtt{CYC}_n (\rho_1 \otimes \ldots \otimes \rho_n)]$ where $\mathtt{CYC}_n$ is a unitary---such as the cycle test~\cite{oszmaniec2024measuring} typically assume that all input states are either fully unknown (e.g., black-box preparations) or need to be physically prepared on demand. We show that if classical descriptions of $m \leq \lfloor n/2 \rfloor$ of these states are available, one can trade the cost of preparing those $m$ states for the cost of measurement and classical post-processing. This substitution  relaxes the circuit requirements for estimating such invariants. The figure shows the case of $n=5$ and $m=2$.}
    \label{fig:conceptual}
\end{figure}

In this work, we advance the results of Ref.~\cite{chiribella2024dimensionindependentweakvalueestimation}. We demonstrate how their circuits can be straightforwardly generalized to estimate Bargmann invariants~\eqref{eq: Bargmann invariant} of any order by replacing the swap test with its generalized version, the \emph{cycle test}~\cite{oszmaniec2024measuring}. We refer to our scheme as a \emph{measurement-enhanced cycle test}, and therefore to the scheme from Ref.~\cite{chiribella2024dimensionindependentweakvalueestimation} as a \emph{measurement-enhanced swap test}. Our results reveal intriguing trade-offs between the ability to estimate higher-order Bargmann invariants, the quantum memory required, the need for controlled unitaries (such as Fredkin gates), and the availability of prior classical information about observables.  Using our scheme, it is possible to act unitarily on $n'$ systems to estimate Bargmann invariants of order $n = n'+ m$ ($m \le n')$, provided that we have a classical description of the $m$ quantum states in order to implement the measurements $A_k = \{\rho_k, \mathbb{1}-\rho_k\}$ for each state $\rho_k$ whose classical description is given, as illustrated in Fig.~\ref{fig:conceptual}. 

The results of Chiribella~et al.~\cite{chiribella2024dimensionindependentweakvalueestimation} demonstrate that the swap test can be employed to estimate $\text{Tr}[\rho_1A\rho_2B]$. Motivated by this, a natural question is whether variations of the \emph{destructive} swap test~\cite{garcia2013swap}---without auxiliary qubits or Fredkin gates---can also be a useful alternative to estimate such multivariate traces in some regimes. Here, we focus on the simplest case of third-order invariants for single-qubit states. We show that it is possible to estimate third-order invariants using circuit variations of a destructive swap test. Nevertheless, we argue that this approach is impractical. While we present a scheme for estimating these invariants given classical state information of one of the three states, the excessive number of required measurements and post-processing renders it useless compared to standard methods or simply performing state tomography. 

Although measurement-enhanced versions of destructive tests may not be useful, we are motivated by the goal of estimating invariants destructively. We therefore discuss how to implement \emph{destructive cycle tests}, which avoid the need for auxiliary qubits to estimate Bargmann invariants, and present a quantum circuit implementation of a destructive 3-cycle test for estimating any third-order Bargmann invariant.

\textbf{Outline.} The remainder of this work is structured as follows. Sec.~\ref{sec: background} presents background on standard tests for estimating Bargmann invariants. It begins with well known schemes for estimating second-order invariants in Sec.~\ref{sec: two-state overlaps}, followed by higher-order invariants in Sec.~\ref{sec: circuits higher order invarints}. Section~\ref{sec: trade off chiribella results} presents the results by Chiribella~et al.~\cite{chiribella2024dimensionindependentweakvalueestimation}, which we later generalize in Sec.~\ref{sec: generalization of chiribella}. We dub our generalized protocol the measurement-enhanced cycle test, encompassing the standard cycle test scheme, the swap test scheme, and the protocol from Ref.~\cite{chiribella2024dimensionindependentweakvalueestimation} as as specific instances.  Section~\ref{sec: measuring third orders with two qubits} investigates if improvements are possible by considering variations of the so-called destructive swap test. In that section we show how to use the destructive swap test to estimate third-order invariants, describe how to implement \emph{destructive} cycle tests, and present a quantum circuit for estimating third-order invariants.  Finally, Sec.~\ref{sec: discussion and outlook} concludes with a discussion of our results, applications, and future directions.

\section{Background}\label{sec: background}

\subsection{Quantum circuits for measuring two-state overlaps}\label{sec: two-state overlaps}

\begin{figure}[t]
    \centering
    \includegraphics[width=0.5\linewidth]{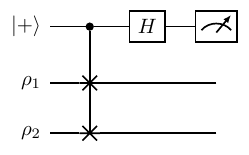}
    \caption{\textbf{Circuit implementing a swap test.} The inputs are two quantum states and an auxiliary qubit system (also known as a control system) that is put in a coherent state $\vert +\rangle \langle + \vert \otimes \rho_1 \otimes \rho_2$. A controlled swap (also known as a Fredkin gate) between the auxiliary qubit and the two states is performed. an Hadamard gate is applied to the auxiliary qubit that is then measured in the $Z$ basis $\{\vert 0\rangle, \vert 1\rangle\}$. The two-state overlap is recovered via the relation $p(0) = (1+\text{Tr}[\rho_1\rho_2])/2$.}
    \label{fig: swap test}
\end{figure}

Starting with quantum circuits used to estimate $\text{Tr}[\rho \sigma]$,  we first review the swap test~\cite{barenco1997stabilization,buhrman2001quantumfingerprinting}. Let us denote by $\mathtt{SWAP}: \mathcal{H} \otimes \mathcal{H} \to \mathcal{H} \otimes \mathcal{H}$  the swap unitary operation such that $\mathtt{SWAP}(\ket u \otimes \ket v) = \ket v \otimes \ket u$, for every $\ket u, \ket v \in \mathcal{H}$. Given a basis $\{\vert i\rangle\}_i$ for $\mathcal{H}$ we can write the swap unitary as 
$$\mathtt{SWAP} = \sum_{i,j} \vert i\rangle \langle j\vert \otimes \vert j \rangle \langle i \vert,$$
from which it is elementary to show that for general density matrices $\rho_1,\rho_2 \in \mathcal{D}(\mathcal{H})$ it holds that

\begin{equation}\label{eq: overlap as average of SWAP on rho1rho2}
    \text{Tr}[\rho_1\rho_2] =  \text{Tr}[\mathtt{SWAP}(\rho_1 \otimes \rho_2)].
\end{equation}
It is then possible to use this relation to propose a quantum circuit for estimating the two-state overlap, known as the swap test~\cite{buhrman2001quantumfingerprinting}. This test performs an Hadamard test~\cite{aaronov2008polynomial} where the unitary $U = \mathtt{SWAP}$. When necessary, to specify that we swap systems $\mathcal{H}_a$ and $\mathcal{H}_b$, we will also write $\mathtt{SWAP}_{a,b}$. The swap test was recently re-discovered by employing machine learning techniques~\cite{schiansky2023demonstration,larocca2022group} and is illustrated in Fig.~\ref{fig: swap test}. For a generalization of this test to multi-qudit systems $\mathcal{H} = (\mathbb{C}^d)^{\otimes n}$ or to infinite dimensional systems, we refer the reader to Refs.~\cite{fujii2003exchange, foulds2024generalising, foulds2021controlledSWAP, foulds2024generalising}.

\begin{figure}[t]
    \centering
    \includegraphics[width=0.5\linewidth]{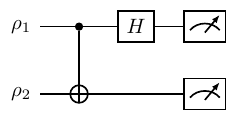}
    \caption{\textbf{Circuit implementing a destructive swap test.} The input state is $\rho_1 \otimes \rho_2$. One then performs a Bell measurement, which is implemented by a CNOT followed by an Hadamard on the first system, and local $Z$ measurements. Assuming each system is a single qubit, the two-state overlap is recovered via the relation $p(1,1) = {(1 - \text{Tr}[\rho_1\rho_2])}/{2}$. For generic multi-qubit systems see Ref.~\cite{bandyopadhyay2023efficient}.}
    \label{fig: destructive SWAP}
\end{figure}

As can be seen in Fig.~\ref{fig: swap test}, to estimate the two-state overlap using the swap test we perform measurements in the auxiliary qubit, leaving the other systems unmeasured. Therefore, the product state $\rho_1 \otimes \rho_2$ is then projected onto another quantum state without being destroyed in the process. Contrastingly, the \emph{destructive} swap test~\cite{garcia2013swap,bandyopadhyay2023efficient,galvao2020quantumandclassical} requires no auxiliary qubit, and uses the fact that the swap operator can be expended in the Bell basis. Therefore, the destructive swap test estimates $\text{Tr}[\rho_1\rho_2]$ by making a Bell measurement on the bipartite system. The quantum circuit implementation is shown in Fig.~\ref{fig: destructive SWAP}. 


For both tests shown in Figs.~\ref{fig: swap test} and~\ref{fig: destructive SWAP} we can imagine that a form of delegated quantum computation is happening, where we do not have complete information of the quantum states that are sent to us by another party. We can perform the quantum computation for this party and return to them the value $\text{Tr}[\rho_1\rho_2]$ we infer without ever knowing the actual states they prepared. 

\subsection{Quantum circuits for measuring Bargmann invariants}\label{sec: circuits higher order invarints}

It is possible to generalize the argument made for the swap test in order to estimate higher-order Bargmann invariants with different implementations of an Hadamard test. We follow Refs.~\cite{oszmaniec2024measuring,quek2024multivariatetrace}. We start noticing that, for every $n$-tuple of quantum states $\pmb\varrho \in \mathcal{D}(\mathcal{H})^n$, we can write \eqref{eq: Bargmann invariant} as
\begin{equation}\label{eq: cycle operator relation}
    \Delta_n(\pmb\varrho) = \text{Tr}[\mathtt{CYC}_n(\rho_1 \otimes \rho_2 \otimes \ldots \otimes \rho_n)],
\end{equation}
where $\mathtt{CYC}_n$ is a unitary associated to the action of a cyclic permutation of $n$ elements 
\begin{equation}\label{eq: cycle permutation}
    (a_1,a_2,a_3\ldots,a_{n-1},a_n) \stackrel{C_n}{\mapsto} (a_2,a_3,\ldots,a_{n-1},a_n,a_1).
\end{equation}
When $n=2$ we have that $\mathtt{CYC}_2 \equiv \mathtt{SWAP}$, and we recover Eq.~\eqref{eq: overlap as average of SWAP on rho1rho2}. For $n\geq 3$, we know that $\Delta_n(\pmb\varrho) \in \mathbb{C}$~\cite{fernandes2024unitaryinvariant}. We can estimate $\Delta_n(\varrho)$ using an Hadamard test for which $U = \mathtt{CYC}_n$. This type of test is known as the \emph{cycle test}~\cite{oszmaniec2024measuring}.

\begin{figure}[t]
    \centering
    \includegraphics[width=\columnwidth]{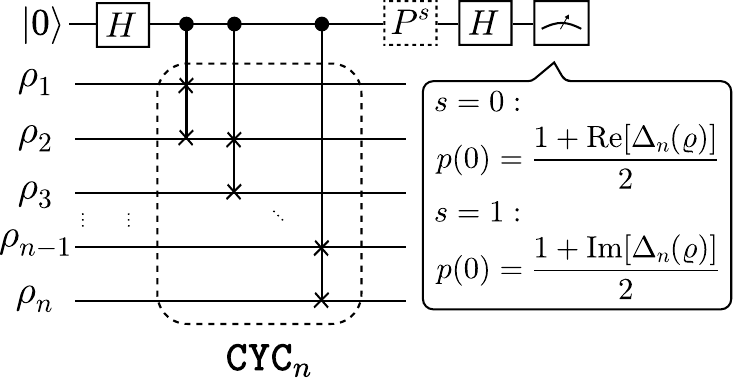}
    \caption{\textbf{Circuit implementing a cycle test.} We show an instance of an Hadamard test, where the unitary $\mathtt{CYC}_n$ is the unitary representation of a cyclic permutation $C_n$ of $n$ symbols. We initialize a quantum memory of $n+1$ systems in a product state. The first wire in the circuit represents a single qubit system. The remaining wires represent systems of dimension $d \geq 2$. After the controlled cycle operation a gate $P^s=\text{diag}(1,i^s)$ is applied to the auxiliary qubit, later measured with the computational basis. When $s = 0$, measuring the auxiliary qubit yields the real part of the Bargmann invariant, while when $s = 1$ the imaginary part.}
    \label{fig: cycle test}
\end{figure}

Different instances of the cycle test~\cite{oszmaniec2024measuring} consider different decompositions of $\mathtt{CYC}_n$ into different elementary gates. A possible choice is via the  sequence of swap unitaries,
\begin{equation}\label{eq: sequence decomposition cyc}
    \mathtt{CYC}_n = \mathtt{SWAP}_{n-1,n}  \ldots  \mathtt{SWAP}_{2,3}\mathtt{SWAP}_{1,2}.
\end{equation}
This leads to the circuit implementation of the cycle test shown in Fig.~\ref{fig: cycle test}. Note that when $n=2$ this reduces to the swap test from Fig.~\ref{fig: cycle test}.

Depending on how the $n$-cycle unitary operator $\mathtt{CYC}_n$ is decomposed into SWAPs (or other unitary gates), each such decomposition yields a different quantum circuit capable of estimating Bargmann invariants. One way of improving the depth of the quantum circuit just described is by using entangled states as auxiliary systems, instead of a single-qubit control system as in Fig.~\ref{fig: cycle test}. We refer to Refs.~\cite{oszmaniec2024measuring,quek2024multivariatetrace} for details. 

So far, all the quantum circuits considered for estimating Bargmann invariants allow for the possibility that all the quantum states in $\pmb\varrho$ are provided by another party. In this way, it is natural that the circuits to estimate Bargmann invariants of $n$ unknown quantum states requires $n$ systems to be available as a quantum memory to be used in the quantum computation. 

However, as recently shown in Ref.~\cite{chiribella2024dimensionindependentweakvalueestimation}, it is possible to estimate third and fourth-order Bargmann invariants using just the swap test circuit from Fig.~\ref{fig: swap test}. They do not explicitly consider this specific use as they focus on the estimation of weak values~\cite{tamir2013introduction,cohen2018determination,wagner2023simple}. Yet, it is trivial to see that their protocol can be used for this purpose. Intriguingly, this allows one to use a quantum circuit with \textit{two} systems (and an additional single qubit auxiliary system) to estimate Bargmann invariants of order \textit{up to four}. This gain is not only in the accessible memory, but also in the number of necessary Fredkin gates required as a standard cycle test to estimate a fourth-order invariant would need three Fredkin gates. The key insight here is that these circuits require classical information on two of the four states. In what follows, we briefly review the results from Ref.~\cite{chiribella2024dimensionindependentweakvalueestimation}.

\subsection{The measurement protocol by Chiribella~et~al.}\label{sec: trade off chiribella results}

\begin{figure}[t]
    \centering
    \includegraphics[width=0.4\linewidth]{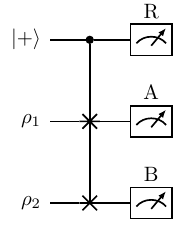}
    \caption{\textbf{Quantum circuit considered by Chiribella~et~al.~\cite{chiribella2024dimensionindependentweakvalueestimation}.} The input state is $\vert +\rangle \langle +\vert \otimes \rho_1 \otimes \rho_2$. The first wire represents a single qubit system, while the remaining wires represent generic single qu\emph{dit} systems. We apply a Fredkin gate between all three systems and perform local measurements. The auxiliary qubit is measured with the positive operator-valued measure (POVM) $R$ from Eq.~\eqref{eq: povm R}. The other systems are measured with generic POVMs $A$ and $B$. We dub this a measurement-enhanced swap test since the measurements of $A,\,B$ allow for the estimation of $\text{Tr}[\rho_1A\rho_2B]$, otherwise inaccessible to the standard swap test---hence the `enhancement' in the order of the accessible multivariate trace.}
    \label{fig: SWAP_test_trade_off}
\end{figure}

The relevant quantum circuit is shown in Fig.~\ref{fig: SWAP_test_trade_off}.  From it, Ref.~\cite{chiribella2024dimensionindependentweakvalueestimation} shows how to estimate $\text{Tr}[\rho_1A\rho_2B]$, where $A$ and $B$ are positive operator-valued measures (POVMs)~\cite{nielsen2010quantum}. Suppose we have access to a tripartite system  $\mathcal{H}_{C} \otimes \mathcal{H}_{S_1} \otimes \mathcal{H}_{S_2} \equiv \mathbb{C}^2 \otimes \mathcal{H} \otimes \mathcal{H}$, where $\mathcal{H}$ is an arbitrary finite dimensional Hilbert space. The system $\mathcal{H}_C = \mathbb{C}^2 $ is our control (or auxiliary) qubit. The algorithm proceeds as follows:

\begin{enumerate}
    \item Prepare a tripartite product state $\vert +\rangle \langle + \vert \otimes \rho_1 \otimes \rho_2$, where $\vert +\rangle = \sfrac{1}{\sqrt{2}}(\vert 0\rangle +\vert 1\rangle) \in \mathbb{C}^2$, while $\rho_1,\rho_2 \in \mathcal{D}(\mathcal{H})$ are generic.
    \item Apply the controlled $\mathtt{SWAP}$ gate, that we denote as $\mathtt{cSWAP}$~\footnote{This is also known as the Fredkin gate.}.
    \item Measure system  $\mathcal{H}_{S_1}$ with a POVM $A=\{P_j\}_j$. 
    \item Measure system $\mathcal{H}_{S_2}$ with a POVM $B = \{Q_k\}_k$.
    \item Measure the auxiliary qubit system $\mathcal{H}_C$ with a POVM $\{R_c\}_{c=0}^3$ given by \begin{align}
    R_0 &= \frac{1}{2}\vert +\rangle \langle + \vert, &R_1& = \frac{1}{2}\vert - \rangle \langle - \vert, \nonumber \\
    R_2 &= \frac{1}{2}\vert +_i\rangle \langle +_i\vert, &R_3& = \frac{1}{2}\vert -_i\rangle \langle -_i \vert,\label{eq: povm R}
\end{align}
where $\vert \pm \rangle = \sfrac{1}{\sqrt{2}}(\vert 0\rangle \pm \vert 1\rangle )$ and $\vert \pm_i \rangle = \sfrac{1}{\sqrt{2}}(\vert 0\rangle \pm i \vert 1\rangle )$.
\end{enumerate}

From this protocol, one obtains the following \textit{joint} distribution:
\begin{align*}
    &p(j,k,c|\rho_1,\rho_2) =\\ &=\text{Tr}[(P_j \otimes Q_k \otimes R_c) \,\mathtt{cSWAP}(\rho_1 \otimes \rho_2 \otimes \vert +\rangle \langle + \vert )\mathtt{cSWAP}^\dagger].
\end{align*}
As shown by Ref.~\cite{chiribella2024dimensionindependentweakvalueestimation}, sampling from this distribution allows us to learn the real and imaginary part of the complex values $$q(j,k|\rho_1,\rho_2) := \text{Tr}[P_j \rho_1 Q_k \rho_2].$$
To see this, let us consider for simplicity the case where $B = \mathbb{1}$. In this case, the joint distribution simplifies to
\begin{align}
    p(j,c|\rho_1,\rho_2) &= \frac{1}{8}\Bigr( \text{Tr}[P_j \rho_1] + \text{Tr}[P_j \rho_2]\nonumber \\
    &+2\theta(1-c)(-1)^c \text{Re}[\text{Tr}[P_j\rho_1\rho_2]]\nonumber\\&-2\theta(c-2)(-1)^c \text{Im}[\text{Tr}[P_j\rho_1\rho_2]]\Bigr), \label{eq: chiribella_etal_old}
\end{align}
where $\theta(t)$ is the Heaviside step function~\cite[Eq.(10), Appendix A]{chiribella2024dimensionindependentweakvalueestimation} and $c \in \{0,1,2,3\}$. We will generalize these results later in Sec.~\ref{sec: generalization of chiribella}.

\begin{figure}[t]
    \centering    \includegraphics[width=\columnwidth]{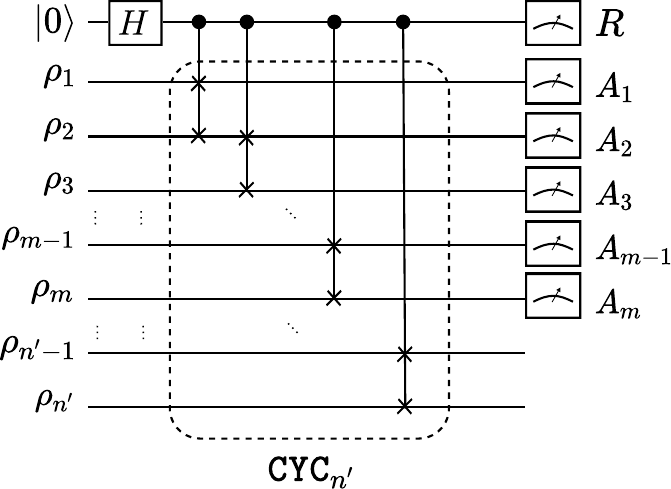}
    \caption{\textbf{Circuit implementing a measurement-enhanced cycle test.} The number of unknown $d$-dimensional quantum states is $n'$ and the number of known states is $m \leq n'$. The order of the estimated multivariate trace is $n = n'+m$. Up to the measurements, the circuit is the same as the one we show in Fig.~\ref{fig: cycle test}. We perform the POVM measurement $R$ on the auxiliary qubit (see Eq.~\eqref{eq: povm R}) and local measurements to $m\leq n'$ systems using the POVMs $A_1,\,\ldots,\,A_m$. }
    \label{fig:measurement-enhanced cycle test}
\end{figure}

Note that in this case estimating this quantity has different demands in terms of prior knowledge of its constituents. While the quantum states $\rho_1,\rho_2 \in \mathcal{D}(\mathcal{H})$ can be completely unknown by who implements the quantum circuit, the operators $P_j,Q_k$ must be known. We shall refer to this quantum algorithm as a measurement-enhanced swap test, as it allows us to estimate more than just two-state overlaps (hence the `enhancement') given that we can perform the measurements $A, \, B$. Specifically, we can see that if $A=\{\rho_3,\mathbb{1}-\rho_3\}$ and $B=\mathbb{1}$ we have that the resulting construction shown above for $p(j,c|\rho_1,\rho_2)$ becomes
\begin{align*}
    p(\rho_3,0) = \frac{1}{8}\Bigr( \text{Tr}[\rho_1\rho_3] + \text{Tr}[\rho_2\rho_3] + 2\text{Re}[\text{Tr}[\rho_1\rho_2\rho_3]]\Bigr),
\end{align*}
where we have let $P_j\equiv \rho_3$, $c=0$, and $p(\rho_3,0) \equiv p(\rho_3,0|\rho_1,\rho_2)$. Hence we conclude that we can learn from this estimation the real part of the third-order Bargmann invariant. Letting $c>2$ allows us to infer the imaginary part of the third-order invariant. In what follows, we will generalize the protocol introduced by Ref.~\cite{chiribella2024dimensionindependentweakvalueestimation}, and then discuss how one can use it to measure multivariate traces of states. 

\section{Measurement-enhanced \\cycle test}\label{sec: generalization of chiribella}

Fix $n \in \mathbb{N}$ to be any integer greater than or equal to $2$. Our goal is to estimate a Bargmann invariant $\text{Tr}[\rho_1 \ldots \rho_n]$ of order $n$. The number of unknown $d$-dimensional quantum states entering the quantum circuit is $n' \leq n$, and we denote them as the tuple $\pmb\varrho = (\rho_1,\ldots,\rho_{n'})$, while the number of quantum states for which we have classical tomographic information is $0 \leq m\leq n'$ such that $n'+ m = n$, and we denote this as the tuple $\widetilde{\pmb{\varrho}} = (\widetilde\rho_1,\ldots,\widetilde\rho_m)$. As we have done previously, we will next introduce how to estimate multivariate traces of states and observables using a generalization of the protocol introduced by Chiribella~et al. and then show how the estimation of Bargmann invariants is a particular instance. In what follows, the $d$-dimensional quantum systems have any finite dimension, and only the auxiliary system is taken to be a single-qubit system.  

We start by considering a generalization of the quantum circuit from Fig.~\ref{fig: SWAP_test_trade_off}. Instead of the $\mathtt{cSWAP}$ considered in Fig.~\ref{fig: SWAP_test_trade_off} we let the controlled unitary to be $\mathtt{cCYC}_n$, i.e. a controlled operation of the unitary $\mathtt{CYC}_n$. We show one possible implementation of such controlled unitary in Fig.~\ref{fig: cycle test}. Letting $\varrho := \rho_1 \otimes \rho_2 \otimes \ldots \otimes \rho_{n'}$ to be the input system on our circuit, the first operation performed is as follows
\begin{equation}
     |+\rangle\langle +| \otimes \varrho \longmapsto \mathtt{cCYC}_{n'}( |+\rangle\langle +| \otimes \varrho ) \mathtt{cCYC}_{n'}^\dagger,
\end{equation}
    where
\begin{equation}
    \mathtt{cCYC}_{n'} = |0\rangle \langle 0| \otimes \mathbb{1}  + |1\rangle \langle 1|\otimes \mathtt{SWAP}_{n', n'-1} \ldots \mathtt{SWAP}_{12}.
\end{equation}
This leaves the entire system in the state 

\begin{align*}
    &\frac{1}{2} \vert 0\rangle \langle 0 \vert \otimes \varrho  + \frac{1}{2} \vert 1\rangle \langle 1 \vert \otimes \mathtt{CYC}_{n'}(\varrho)+\\
    &+ \frac{1}{2}\Bigr( \vert 0\rangle \langle 1 \vert \otimes\sum_{\mathbf{a}}p_{\mathbf{a}}\vert a_2\rangle \langle a_1 \vert \otimes \ldots \otimes \vert a_1\rangle \langle a_{n'} \vert + \mathrm{h.\,c.}\Bigr),
\end{align*}
where $\mathbf{a} = (a_1,a_2,\ldots,a_{n'})$ is a multi-index label which we use as $\sum_{\mathbf{a}} \equiv \sum_{a_1,a_2,\ldots,a_{n'}}$. Moreover $p_{\mathbf{a}} = p_{a_1}^1p_{a_2}^2\ldots p_{a_{n'}}^{n'}$ is the product statistics, and we have written each state $\rho_i$ in terms of \emph{some}---not necessarily known---convex combination of pure quantum states  $\rho_i = \sum_{a_i}p_{a_i}^i \vert a_i\rangle \langle a_i \vert.$ We have also used 
\begin{align*}
\mathtt{CYC}_{n'}(\vert a_1,a_2,\ldots,a_{n'}\rangle) &= \vert a_{2},\ldots,a_{n'},a_1\rangle.
\end{align*}

If we stop here, measuring the auxiliary qubit in the $X$ ($Y$) basis  performs the cycle test operation for estimating the real (imaginary) part of a $n'$-th order Bargmann invariants~\cite{oszmaniec2024measuring,wagner2024quantum}. Now let us assume that not only the control qubit is measured on the $X$ basis, but also {the first $m \leq n'$ systems}, for some fixed value $m$,  are measured with respect to POVMs $A_1 = \{P_{j_1}\}_{j_1}$, $\ldots$, $A_m = \{P_{j_m}\}_{j_m}$. This yields a joint probability distribution
\begin{widetext}
\begin{eqnarray*}
    p(j_1, \ldots, j_m, \pm) &=& \frac{1}{4} \Bigl( \prod_{i=1}^m \operatorname{Tr}[P_{j_i} \rho_i] + \prod_{i=1}^m  \operatorname{Tr}[P_{j_i} \rho_{i+1}] \pm\sum_{\mathbf{a}} p_{\mathbf{a}} \langle a_2| P_{j_1} | a_1\rangle \langle a_3| P_{j_2} | a_2\rangle \ldots  \langle a_{m+1}| P_{j_m} | a_m\rangle \times \\
    &\times& \langle a_{m+2}| a_{m+1}\rangle \ldots \langle a_{n'}| a_{n'-1}\rangle \langle a_1| a_{n'}\rangle \pm \mathrm{c.\, c.} \Bigr) \\
    &=& \frac{1}{4} \Bigl( \prod_{i=1}^m \operatorname{Tr}[P_{j_i} \rho_i] + \prod_{i=1}^m  \operatorname{Tr}[P_{j_i} \rho_{i+1}] \pm \sum_{\mathbf{a}}p_{\mathbf{a}}\langle a_1| a_{n'}\rangle \langle a_{n'}| a_{n'-1}\rangle \ldots \langle a_{m+2}| a_{m+1} \rangle \times \\
    &\times&  \langle a_{m+1}| P_{j_m} | a_m\rangle \ldots  \langle a_3| P_{j_2} | a_2\rangle \langle a_2| P_{j_1} |a_1\rangle  \pm \mathrm{c.\,c.} \Bigr) \\
    &=& \frac{1}{4} \Bigl( \prod_{i=1}^m \operatorname{Tr}[P_{j_i} \rho_i] + \prod_{i=1}^m  \operatorname{Tr}[P_{j_i} \rho_{i+1}] \pm  2\operatorname{Re} [\operatorname{Tr}(\rho_{n'} \ldots \rho_{m+2}\rho_{m+1}P_{j_m}\rho_m P_{j_{m-1}}\rho_{m-1}\ldots P_{j_2}\rho_2 P_{j_1}\rho_1)] \Bigr).
\end{eqnarray*}
\end{widetext}

To simplify the notation we write $\mathbf{j} = (j_1,j_2,\ldots,j_m)$, and we also denote  
\begin{equation}\label{eq: square states and effects}
    \square_{n'+m}(\mathbf{j},\pmb\varrho) := \operatorname{Tr}(\rho_{n'} \ldots \rho_{m+1}P_{j_m}\rho_m \ldots  P_{j_1}\rho_1).
\end{equation}
Note that the above is \emph{not} a Bargmann invariant, since each $P_{j_m}$ is \emph{not} necessarily a quantum state, but a generic POVM element (hence the different notation of using $\square$ instead of $\Delta$)~\footnote{Note that for any $\mathbf{j}$ and any $\pmb\varrho$ we have that there exists a value $c \in \mathbb{R}$ and some Bargmann invariant $\Delta_{n}$ such that $\square_{n}(\mathbf{j},\pmb\varrho) = c \cdot \Delta_{n}$. In other words, $\square_{n}(\mathbf{j},\pmb\varrho)$ is always just a scaling of some Bargmann invariant as given by Eq.~\eqref{eq: Bargmann invariant}. See also the discussion from Ref.~\cite{wagner2024quantum} on the differences between extended Kirkwood--Dirac quasiprobability distributions and Bargmann invariants.}. 

Instead of performing an $X$ measurement of the auxiliary qubit, we can perform random measurements with respect to both $X$ and $Y$ bases on the auxiliary (control) qubit, using the four outcome POVM $R$ from Eq.~\eqref{eq: povm R}. In this case, the joint probability distribution becomes:
\begin{align*}
    p(\mathbf{j},c) &= \frac{1}{8} \Bigl( \prod_{i=1}^m \operatorname{Tr}[P_{j_i} \rho_i] + \prod_{i=1}^m \operatorname{Tr}[P_{j_i} \rho_{i+1}]\\
    &+ 2\theta  (1-c)  (-1)^c {\rm Re}  \left[ \square_{n'+m}(\mathbf{j},\pmb\varrho)\right]\\&- 2\theta (c-2) (-1)^c  {\rm Im}  \left[ \square_{n'+m}(\mathbf{j},\pmb\varrho)\right]\Bigr),
\end{align*}
where we have used $\theta(t)$ to denote the Heaviside step function, and $c \in \{0,1,2,3\}$ denotes the outcomes from the POVM measurement $R = \{R_c\}_c$ of the auxiliary qubit. The above generalizes the construction considered in Ref.~\cite{chiribella2024dimensionindependentweakvalueestimation}, that was reviewed in Sec.~\ref{sec: trade off chiribella results}. Specifically, we recover Eq.~\eqref{eq: chiribella_etal_old} by making $m=1$ and $j_1\equiv j$. 

If we have $m$ observables $A_m$ with the corresponding POVM decompositions 
\begin{equation}\label{eq: POVM observable}
A_m = \sum_j x_{j_m} P_{j_m},
\end{equation}
we can define a random variable $\tilde{X}$ that takes values
\begin{align}
\widetilde x_{j_1, \ldots, j_{m}, c}   :  =  2  x_{j_1} \ldots x_{j_{m}} \,  (-1)^c  \, \Big[   \, \theta(1-c) - i \theta (c-2)\Big] \label{eq:ranvar}
\end{align}
with respect to the probability distribution $p(\mathbf{j}, c)$. Note that the values $x_{j_i}$ are given by Eq.~\eqref{eq: POVM observable}. The expectation value of $\tilde{X}$ returns the value of $\square_{n'+m}({\mathbf{j}},\pmb\varrho)$ since
\begin{align}
    &\mathbb{E}[\tilde{X}] 
    =\nonumber\\
    &=\sum_{j_1, \ldots, j_{m}, c} 2  \prod_{i=1}^{m} x_{j_i} \,  (-1)^c  \, \Big[   \, \theta(1-c) - i \theta (c-2)\Big] p(\mathbf{j}, c) \label{eq: part_1_t}\\
    &= \sum_{j_1, \ldots, j_{m}} x_{j_1} \ldots x_{j_{m}} \Bigl( {\rm Re}[\square_{n}({\mathbf{j}},\pmb\varrho)] + i  {\rm Im}[\square_{n}({\mathbf{j}},\pmb\varrho)]\Bigr) \label{eq: part_2_t}\\
    &=  \sum_{j_1, \ldots, j_{m}} x_{j_1} \ldots x_{j_{m}}\square_{n}({\mathbf{j}},\pmb\varrho). \label{eq: conclusion_expectation}
\end{align}
Above, we have used that $n = n'+ m$. The terms having products of traces $\text{Tr}[P_{j_i}\rho_i]$ cancel out going from Eq.~\eqref{eq: part_1_t} to Eq.~\eqref{eq: part_2_t} due to the sum in $c$, since we get a sum of contributions   $(-1)^0 + (-1)^1 + i ((-1)^2 + (-1)^3) = 1-1+i-i =0$. This means that Eq.~\eqref{eq: conclusion_expectation} guarantees we can estimate a quantity $\square_{n'+m}({\mathbf{j}},\pmb\varrho)$ given by Eq.~\eqref{eq: square states and effects} by sampling from the distribution generated by performing controlled SWAPs on $n'$ systems and measuring observables $\{A_{i}\}_{i=1}^m$ on $m \leq n'$ of these systems. We remark that (see for instance  Refs.~\cite{chiribella2024dimensionindependentweakvalueestimation,wagner2024quantum,quek2024multivariatetrace}) estimating such traces is exponentially better (in sample complexity) than performing quantum state tomography of all unknown quantum states $n'$.  
 
We now conclude by applying our test to the estimation of Bargmann invariants. We do so by restricting the POVMs to be such that $A_i = \{\widetilde\rho_i,\mathbb{1}-\widetilde\rho_i\}$.  In this case, selecting the specific case where the labels satisfy $(P_{j_1},\ldots,P_{j_m}) = (\widetilde \rho_1,\ldots,\widetilde \rho_m) \equiv \widetilde {\pmb\varrho}$ we end up with 
\begin{equation}
    \Delta_{n'+m}(\widetilde{\pmb\varrho},\pmb\varrho) = \Tr[\rho_{n'}\ldots \rho_{m+2}\rho_{m+1}\widetilde{\rho}_m\rho_m\ldots\widetilde{\rho}_1\rho_1].
\end{equation}
This means that any  $n$-th order Bargmann invariant can be estimated by using a cyclic shift on $n' \leq n$ quantum states (hence, $n'-1$ controlled SWAPs) and measurements on $m \leq n'$ systems where $n=n'+m$. Alternatively, at least $\lceil n/2 \rceil$ systems and one auxiliary qubit, and hence $\lceil n/2 \rceil-1$ controlled SWAPs, are necessary to estimate $n$-th order Bargmann invariants using our sampling protocol and the decomposition of $\mathtt{CYC}_n$ as by Eq.~\eqref{eq: sequence decomposition cyc}. Hence, the sampling protocol requires only \textit{half} of systems and controlled SWAPs compared with the original cycle test protocol of Refs.~\cite{oszmaniec2024measuring,wagner2024quantum}.

\section{Destructive tests for estimating Bargmann invariants}\label{sec: measuring third orders with two qubits}

So far, we have considered the situation where we have generalized the cycle test to include classical information of observables and investigated how this allows us to access information of higher order multivariate traces. It is natural to ask whether the \emph{destructive} swap test can also be generalized to estimate invariants \eqref{eq: Bargmann invariant} of order larger than $n=2$, provided we have access to prior classical tomographic information of some of the states. 

While in the last section the states in $\pmb{\rho}$ where taken to be generic quantum states on a finite-dimensional Hilbert space, i.e. $\pmb{\rho} \in \mathcal{D}(\mathbb{C}^d)^n$, in this section we restrict to the case where $d=2$, i.e., every state in the tuple is a single-qubit quantum state.  

We start by discussing how a third-order Bargmann invariant can be estimated under certain assumptions on prior knowledge on the corresponding states using instances of a deterministic swap test. First, we notice that \textit{every} third-order Bargmann invariant is quantum realizable by single qubit pure quantum states~\cite{fernandes2024unitaryinvariant}. Therefore, without loss of generality, we consider $|\psi_1\rangle, | \psi_2 \rangle, | \widetilde\psi_3 \rangle \in \mathbb{C}^2$ to be pure single-qubit states, where $|\psi_1\rangle$ and $| \psi_2 \rangle$ are not necessarily known. The corresponding third-order Bargmann invariant is given by: 
\begin{equation}\label{eq:3invPure}
\Delta_3(\psi_1, \psi_2, \widetilde\psi_3) = \langle \psi_1| \psi_2\rangle \langle \psi_2 | \widetilde\psi_3 \rangle \langle \widetilde\psi_3|\psi_1 \rangle,
\end{equation}
where we use the notation $\psi_i:=|\psi_i\rangle\langle\psi_i|$ for pure states in the argument of Bargmann invariants. 

On the other hand, taking into account the unitary-invariance of Bargmann invariants~\cite{galvao2020quantumandclassical}, for any triplet $\vert \psi_1\rangle, \vert \psi_2 \rangle, \vert \tilde\psi_3 \rangle \in \mathbb{C}^2$ we can always find some unitary $U$ such that $U \vert \tilde\psi_3 \rangle = \vert 0\rangle$ using the available tomographic classical information of $\vert \tilde{\psi}_3\rangle $. In this case, 
\begin{align*}
    \Delta_3(\psi_1, \psi_2, \widetilde\psi_3) &= \langle \psi_1|U^\dagger U\vert \psi_2 \rangle \langle \psi_2 | U^\dagger |0\rangle \langle 0 | U |\psi_1 \rangle \\&:= \langle \psi_U|\phi_U\rangle \langle \phi_U \vert 0\rangle \langle 0 | \psi_U \rangle \\
    &=\Delta_3(\psi_U, \phi_U, 0),  
\end{align*}
where $\vert\psi_U\rangle=U\vert \psi_1\rangle$, and $\vert\phi_U\rangle = U\vert \psi_2 \rangle$. Therefore, given access to tomographic classical information of $\vert \tilde{\psi}_3\rangle $, estimation of a generic Bargmann invariant \eqref{eq:3invPure} is equivalent to estimation of Bargmann invariant on the triplet $U\vert \psi_1\rangle$, $U\vert \psi_2 \rangle$, and $\vert 0 \rangle$.

\begin{figure}[t!]
    \centering
    \includegraphics[width=0.75\linewidth]{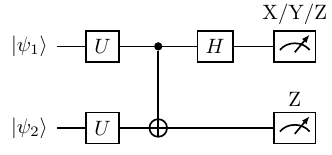}
    \caption{\textbf{Circuit implementing a modified destructive swap test for the estimation of third order Bargmann invariants.} The input state is $|\psi_1 \rangle \otimes |\psi_2\rangle$. One then applies a unitary operator $U$ defined as $U|\widetilde\psi_3\rangle = |0\rangle$, followed by a combination of CNOT and Hadamard gates similarly to destructive swap test. Finally, the second qubit undergoes a $Z$ measurement, while the first qubit is measured in $X$-, $Y$-, or $Z$-basis. Assuming each system is a single qubit, the third order Bargmann invariant is given by Eq.~\eqref{eq:3bargInvKnown}. 
    } 
    \label{fig: chi destructive SWAP}
\end{figure}

Notice that the projector $ | 0 \rangle \langle 0|$ can be decomposed into operators $\frac{1}{2}(\mathbb{1} \pm Z)$, where $Z$ is the Z-Pauli operator. Therefore, we can write 
\begin{eqnarray}
    \nonumber \Delta_3(\psi_1, \psi_2, \widetilde\psi_3) &=& \frac{1}{2}\Bigl(\Delta_2(\psi_U, \phi_U) + \chi(\psi_U, \phi_U) \Bigr),
\end{eqnarray}
where $\chi(\psi_U, \phi_U) := \langle \psi_U | \phi_U \rangle \langle \phi_U | Z | \psi_U \rangle$. Since $\Delta_2(\psi_U, \phi_U)$ can be estimated via the destructive swap test discussed in Sec.~\ref{sec: two-state overlaps}, the core challenge lies in estimation of $\chi(\psi_U, \phi_U)$. This can be achieved using the quantum circuit shown in Fig.~\ref{fig: destructive SWAP}, where, in contrast to estimation of second order Bargmann invariant, only the second qubit is measured in computational basis, while the first qubit is measured in $X$- and $Y$-basis in order to obtain its real and imaginary parts, respectively. In order to proof this, we take into account that, decomposing the states as | $\psi_U \rangle = a |0\rangle + a' |1\rangle$ and $| \phi_U \rangle = b |0\rangle + b' |1\rangle$, with $|a|^2 + |a'|^2 = 1$ and $|b|^2 + |b'|^2 = 1$, we can write
\begin{equation}
    \chi(\psi_U, \phi_U) = |a|^2 |b|^2 - |a'|^2 |b'|^2 + 2i \mathrm{Im}[ab^*a'^*b'].
\end{equation}
On the other hand, the circuit in Fig.~\ref{fig: destructive SWAP} produces a state
\begin{align}
    &|\Psi\rangle = \frac{1}{\sqrt{2}}\Biggl( |0\rangle  \Bigl((ab + a'b')|0\rangle + (ab' + a'b)|1\rangle\Bigr) \nonumber \\&+ |1\rangle  \Bigl((ab - a'b')|0\rangle + (ab' - a'b)|1\rangle\Bigr)\Biggr).\label{eq:hCnotOutput}
\end{align}
before performance of local measurements. Then, a straightforward calculation shows that a measurement of the first qubit in $X$-basis and the second qubit in computational basis yields the probabilities
\begin{eqnarray}
    p(+,0) &=& |a|^2|b|^2, \\
    p(-,0) &=& |a'|^2|b'|^2,
\end{eqnarray}
while measuring the first qubit in the $Y$-basis provides
\begin{equation}
    p(\pm_i,1) = |a|^2 |b'|^2 + |a'|^2 |b|^2 \mp \mathrm{Im}[ab^*a'^*b'],
\end{equation}
where $p(\pm,0)$ and $p(\pm_i,1)$ are the corresponding probabilities of finding the qubits in states $|\pm\rangle\otimes|0\rangle$ and $|\pm_i\rangle\otimes|1\rangle$, respectively. Therefore, we conclude that
\begin{equation}
    \chi(\psi_U, \phi_U) = p(+,0) - p(-,0) - i\Bigl(p(+_i,1) - p(-_i,1) \Bigr),
\end{equation}
In turn, measuring both qubits in computational basis recovers destructive swap test discussed in Sec.~\ref{sec: two-state overlaps}. Therefore, taking into account unitary invariance of Bargmann invariants, we conclude that the third order Bargmann invariant can be estimated by
\begin{align}
    \nonumber \Delta_{3}(\psi_1, \psi_2, \widetilde \psi_3) &= \frac{1}{2}\Bigl( 1 - 2p(1,1) + p(+,0) \\&- p(-,0) - i\Bigl(p(+_i,1) - p(-_i,1) \Bigr)\Bigr). \label{eq:3bargInvKnown}
\end{align}

We observe that a \emph{destructive cycle test} can be introduced to estimate Bargmann invariants and to explore potential measurement-enhanced variants. Although the analysis of the destructive swap test suggests that such enhancements may offer limited benefit, destructive cycle tests---extending the swap test case from Fig.~\ref{fig: destructive SWAP}---may still be of independent interest. Motivated by this, we now show that, in principle, a generic destructive cycle test can be described in a conceptually straightforward manner. Due to Eq.~\eqref{eq: cycle operator relation}, $n$-th order Bargmann invariants for a tuple of states $\pmb\varrho= (\rho_1, \ldots, \rho_n)$ can be seen as expectation values of the $n$-cycle unitary operator $\mathtt{CYC}_n$ with respect to the state $\varrho = \rho_1 \otimes \ldots \otimes \rho_n$,
\begin{equation}\label{eq:bargInvCycN}
    \Delta_n(\rho_1,\ldots,\rho_n) = \langle \mathtt{CYC}_n\rangle_\varrho.
\end{equation}
Therefore, a destructive cycle test can be constructed using diagonalization of the unitary operator $\mathtt{CYC}_n$ with respect to an entangled basis. 

Let us consider the case of qubit states $\rho_1,\ldots,\rho_n \in \mathcal{D}(\mathbb{C}^2)$ and the corresponding total Hilbert space $\mathcal{H} = (\mathbb{C}^2)^{\otimes n}$. The former can be spanned by computational basis states that form a set $\mathcal{S}_n=\{|\mathbf{x}\rangle\}_{\mathbf{x} \in \{0,1\}^n}$, where $\mathbf{x} \in \{0,1\}^n$ is a string of $n$ bits. Then $\mathcal{H}$ can be decomposed into $n+1$ orthogonal subspaces
\begin{equation}\label{eq:decompHW}
\mathcal{H} = \bigoplus_{k=0}^n \mathcal{H}_{\mathrm{HW}}^{(k)},
\end{equation}
where $\mathcal{H}_{\mathrm{HW}}^{(k)}$ are spanned by sets $\mathcal{S}_n^{(k)} \subseteq \mathcal{S}_n$ of ${n\choose k}$ states $|\mathbf{x}_k\rangle$ corresponding to bit strings $\mathbf{x}_k$ of fixed Hamming weight $k$, i.e., bit strings with $k$ 1's. For example, for $n=3$, this decomposition is given by
\begin{align*}
    \mathcal{S}_3^{(0)} &= \{\vert 000\rangle \},\\
    \mathcal{S}_3^{(1)} &= \{\vert 001\rangle, \vert 010\rangle, \vert 100\rangle   \},\\
    \mathcal{S}_3^{(2)} &= \{\vert 110\rangle, \vert 101\rangle ,\vert 011\rangle  \},\\
    \mathcal{S}_3^{(3)} &= \{\vert 111\rangle \}.
\end{align*}
The operator $\mathtt{CYC}_n$ \emph{preserves} the Hamming weight of computational basis states, hence, it acts invariantly on each subset $\mathcal{S}_n^{(k)}$ and corresponding subspace of decomposition in  Eq.~\eqref{eq:decompHW}, so that $
\mathtt{CYC}_n(\mathcal{H}_{\mathrm{HW}}^{(k)}) \subseteq \mathcal{H}_{\mathrm{HW}}^{(k)}$. Moreover, each subset $\mathcal{S}_n^{(k)}$ can be further decomposed into $c_k$ \emph{cyclic orbits} $\{\mathcal{O}^{(k,m)}_n\}_{m=0}^{c_k}$ with respect to the action of $\mathtt{CYC}_n$. These are equivalence classes $\mathcal{S}_n/\sim_\mathsf{cyc}$ under an equivalence relation which, for any $|\mathbf{x}_1\rangle, |\mathbf{x}_2\rangle \in \mathcal{S}_n$, is defined as follows: 
\begin{equation}
    \nonumber |\mathbf{x}_1\rangle \sim_\mathsf{cyc} |\mathbf{x}_2\rangle \Leftrightarrow \exists j \in \{0,\ldots,n-1\}, \; |\mathbf{x}_1\rangle = \mathtt{CYC}_n^j ,|\mathbf{x}_2\rangle.
\end{equation}
 Therefore, each cyclic orbit can be defined as
\begin{eqnarray}
    \mathcal{O}^{(k,m)}_n &=& [\vert \mathbf{x} \rangle ]_{\sim_\mathsf{cyc}} = \left\{ \mathtt{CYC}_n^j \vert \mathbf{x}\rangle\right\}_{j = 0}^{r^{(k,m)}_n-1},
\end{eqnarray}
where $r^{(k,m)}_n$ is the size of the cyclic orbit $\mathcal{O}^{(k,m)}_n$, i.e., the minimal period such that $\mathtt{CYC}_n^{r^{(k,m)}_n} \ket{\mathbf x} = \ket{\mathbf x}$. Therefore, each $\mathcal{H}_{\mathrm{HW}}^{(k)}$ can be further decomposed into $r^{(k,m)}_n$-dimensional invariant subspaces spanned by the corresponding cyclic orbits. 

For the sake of illustration, while in the case $n=3$ the cyclic orbits $\mathcal{O}^{(k,m)}_n$ coincide with entire subsets $\mathcal{S}_3^{(k)}$, the case $n=4$ provides a non-trivial decomposition of $\mathcal{S}_4^{(k)}$. In particular, the subset $\mathcal{S}_4^{(2)} = \{|0011\rangle ,|0101\rangle,|0110\rangle,|1001\rangle,|1010\rangle,|1100\rangle\}$ can be split into two cyclic orbits 
\begin{align*}
    \mathcal{O}_4^{(2,0)} &= \{\vert 0011\rangle, \vert 0110\rangle, \vert 1100\rangle, \vert 1001\rangle \},\\
    \mathcal{O}_4^{(2,1)} &= \{\vert 1010\rangle,\vert 0101\rangle \},
\end{align*}
of periods $r^{(2,0)}_4 = 4$ and $r^{(2,1)}_4 = 2$, respectively.

The $n$-cycle operator $\mathtt{CYC}_n$ can be diagonalized in the basis obtained by using the quantum Fourier transform unitary on each cyclic orbit, defining hence the vectors 
\[
\ket{\psi^{(k,m)}_{n,\ell}} = \frac{1}{\sqrt{r^{(k,m)}_n}} \sum_{j=0}^{r^{(k,m)}_n-1} (\omega^{(k,m)}_{n,\ell})^j \ket{\mathbf{x}_{n,j}^{(k,m)}},
\]
where $\{\ket{\mathbf{x}_j^{(k,m)}}\}_{j=0}^{r^{(k,m)}_n-1} = \mathcal{O}^{(k,m)}_n$, and $\omega^{(k,m)}_{n,\ell}$ is an $r^{(k,m)}_n$-th root of unity $
\omega^{(k,m)}_{n,\ell} = e^{2\pi i\ell/r^{(k,m)}_n}$, with $\ell=0,\ldots,r^{(k,m)}_n-1$, associated to the size of the corresponding cyclic orbit $\mathcal{O}^{(k,m)}_n$. Then, by construction, these states are eigenstates of the $\mathtt{CYC}_n$ with the corresponding eigenvalues $\omega_\ell$,
\[
\mathtt{CYC}_n \ket{\psi^{(k,m)}_{n,\ell}} = \omega^{(k,m)}_{n,\ell} \ket{\psi^{(k,m)}_{n,\ell}},
\]
for all $\ell$. The complete set of vectors $\{ \ket{\psi^{(k,m)}_{n,\ell}}\}_{\ell,k,m}$ provided by all cyclic orbits describes an orthonormal eigenbasis of $\mathcal{H} = (\mathbb{C}^{2})^{\otimes n}$ with respect to $\mathtt{CYC}_n$. Therefore, from Eq.~\eqref{eq:bargInvCycN}, $n$-th order Bargmann invariant can be estimated by preparing $n$ qubits in the product state $\varrho = \rho_1 \otimes \ldots \otimes \rho_n$ and performing the measurements with respect to this eigenbasis:
\begin{equation}
    \Delta_n(\rho_1,\ldots,\rho_n) = \sum_{\ell,k,m} \omega^{(k,m)}_{n,\ell} \text{Tr}\left[\varrho \vert \psi_{n,\ell}^{(k,m)}\rangle \langle \psi_{n,\ell}^{(k,m)}\vert \right ].
\end{equation}

\begin{figure}[t]
    \centering
    
        \includegraphics[width=\columnwidth]{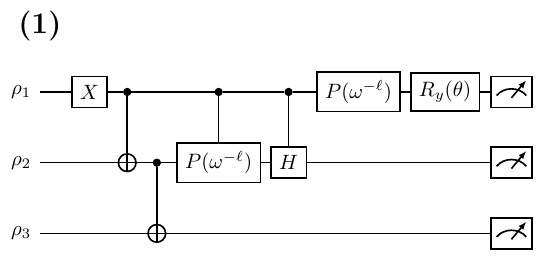}
    
        \includegraphics[width=\columnwidth]{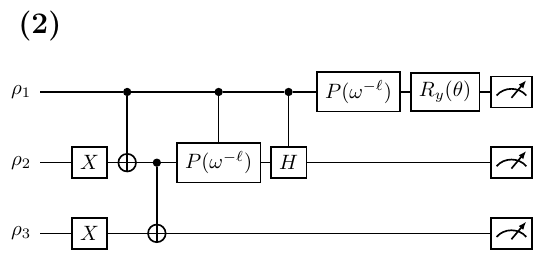}
    
    \caption{\textbf{Circuits for implementing a destructive $3$-cycle test.} The input state is $\rho_1 \otimes \rho_2 \otimes \rho_3$. The circuits start with application of a Pauli $X$-gate that depends on the circuit index \textbf{(k)}: in the circuit \textbf{(1)}, it is applied to the first qubit, while the circuit \textbf{(2)} applies it to the second and third qubits. They are followed by two CNOT gates, controlled phase gate, and controlled Hadamard gate. The circuits are concluded by application of a phase gate and a $Y$-rotation gate on angle $\theta$ on the first qubit, where $\theta = -2\mathrm{arccos}(1/\sqrt{3})$, so that $R_y(\theta) = \sfrac{1}{\sqrt{3}}(\mathbb{1}+i\sqrt{2}\,Y)$. The phase gate is defined as $P(\omega^{-\ell}) = |0\rangle\langle 0| + \omega^{-\ell}|1\rangle \langle 1|$, corresponding to estimation of probabilities $p_{\ell}^{(k)}$ in \eqref{eq:destrCycTestProb}, and $\omega = e^{2\pi i / 3}$. Finally, all qubit get locally measured in the computational basis, providing the necessary statistics for Eq.~\eqref{eq: Bargmann invariant destructive cyc test}.}
    \label{fig:Destructive_cycle_test} 
\end{figure}

We refer to this method of estimating Bargmann invariants as the \emph{destructive cycle test}. For concreteness, let us restrict our attention to the case $n=3$. The operator \( \mathtt{CYC}_3 \) is diagonal in the orthonormal basis composed of two product states
    \[
    \ket{\psi_{3,0}^{(0)}} = \ket{000}, \quad \ket{\psi_{3,0}^{(3)}} = \ket{111},
    \]
and six $W$-like states with complex phases
    \begin{align*}
    \ket{\psi^{(1)}_{3,0}} &= \frac{1}{\sqrt{3}}(\ket{001} + \ket{010} + \ket{100}), \\
    \ket{\psi^{(2)}_{3,0}} &= \frac{1}{\sqrt{3}}(\ket{110} + \ket{101} + \ket{011}), \\
    \ket{\psi^{(1)}_{3,1}} &= \frac{1}{\sqrt{3}}(\omega^2\ket{001} + \omega\ket{010} + \ket{100}), \\
    \ket{\psi^{(2)}_{3,1}} &= \frac{1}{\sqrt{3}}(\omega^2\ket{110} + \omega\ket{101} + \ket{011}), \\
    \ket{\psi^{(1)}_{3,2}} &= \frac{1}{\sqrt{3}}(\omega\ket{001} + \omega^2\ket{010} + \ket{100}), \\
    \ket{\psi^{(2)}_{3,2}} &= \frac{1}{\sqrt{3}}(\omega\ket{110} + \omega^2\ket{101} + \ket{011}),
    \end{align*}
where \( \omega = e^{2\pi i/3} \) is a cube root of unity, and the index $m$ is omitted since, for $n=3$, each subset $\mathcal{S}_n^{(k)}$ has a unique cyclic orbit. Therefore,
\begin{align*}
\mathtt{CYC}_3 &= \sum_{k=0}^3 \Pi_{3,0}^{(k)}
+ \omega (\Pi_{3,1}^{(1)} + \Pi_{3,1}^{(2)}) + \omega^2 (\Pi_{3,2}^{(1)} + \Pi_{3,2}^{(2)}),
\end{align*}
where $\Pi_{3,\ell}^{(k)} = \vert \psi^{(k)}_{3,\ell} \rangle \langle \psi^{(k)}_{3,\ell} |$.~This means that $\Delta_3(\rho_1,\rho_2,\rho_3)$ can be estimated by preparing three qubits in the product quantum state \( \varrho = \rho_1 \otimes \rho_2 \otimes \rho_3 \) and performing a projective measurement in this basis, collecting the corresponding probabilities,
\begin{eqnarray}
\nonumber \Delta_3(\rho_1,\rho_2,\rho_3) &=& 1 - (1 - \omega)(p_{1}^{(1)} + p_{1}^{(2)}) \\
&-& (1 - \omega^2)(p_{2}^{(1)} + p_{2}^{(2)}),\label{eq: Bargmann invariant destructive cyc test} 
\end{eqnarray}
where $p_{\ell}^{(k)} = \text{Tr}[(\rho_1 \otimes \rho_2 \otimes \rho_3) \Pi_{3,\ell}^{(k)}]$. 
Similarly to destructive swap test, due to the cyclic property of trace, measurement in the eigenbasis of $\mathtt{CYC}_3$ can be substituted by local measurements in computational basis using unitary transformations defined via $|\psi_{3,\ell}^{(k)}\rangle = U_{3,\ell}^{(k)}|000\rangle$,
\begin{equation}\label{eq:destrCycTestProb} 
    p_{\ell}^{(k)} = \text{Tr}[U_{3,\ell}^{(k)\dagger}(\rho_1 \otimes \rho_2 \otimes \rho_3)U_{3,\ell}^{(k)} \vert 000\rangle \langle 000 \vert ] := \tilde{p}_{\ell}^{(k)}(0,0,0),
\end{equation}
which can be realized via quantum circuits given in Fig.~\ref{fig:Destructive_cycle_test}. Therefore, third order Bargmann invariant can be estimated in a \emph{destructive} $3$-cycle test by collecting probabilities $\tilde{p}_{\ell}^{(k)}(0,0,0)$ of obtaining the outcomes $0$ in all local measurements in computational basis in circuit \textbf{(k)} with the choice of phase $\omega^{-\ell}$ in the corresponding phase gates. 

To the best of our knowledge, a destructive cycle test had not been introduced before. Nonetheless, Ref.~\cite{reascos2023quantumcircuits} had previously observed that $\langle \mathtt{CYC}_3 \rangle_\varrho$ can be written as the average of two measurements, arriving at a conclusion that is similar in spirit to ours.

\section{Discussion and outlook}\label{sec: discussion and outlook}

In this work, we have generalized the quantum circuits proposed by Chiribella~et al.~\cite{chiribella2024dimensionindependentweakvalueestimation} to estimate multivariate traces of quantum states, introducing a \emph{measurement-enhanced cycle test}. By leveraging prior classical information about observables, our protocol reduces the quantum resources required to estimate Bargmann invariants of order \(n\), cutting the number of qubits and Fredkin gates by up to half compared to the standard cycle test. This trade-off—exchanging classical descriptions of \(m\) states for reduced quantum hardware demands—makes our protocol particularly suitable for near-term quantum devices, where qubit counts and gate fidelities are limiting factors.

The ability to estimate higher-order invariants with limited quantum resources opens new opportunities for practical applications. For instance, in \textit{extended Kirkwood-Dirac (KD) quasiprobability distributions}~\cite{arvidssonshukur2024properties,lostaglio2023kirkwooddirac,wagner2024quantum,gherardini2024quasiprobabilities} the defining bases are often classically known. Our protocol allows for the efficient estimation of extended KD distributions even when only a small quantum processor is available. As another example, in the context of \textit{sequential weak values}~\cite{mitchison2007sequential,diosi2016structural,georgiev2018probing} one estimates high-order multivariate traces of states where two states---namely the pre- and post-selected states---are assumed to be classically known. Yet another example might be that of estimating geometric phases. It is conceivable that one's goal is to estimate the geometric phase with respect to a closed path of $n=n'+m$ points in projective space, where $n'$ of these are unknown vector states while $m$ are classically known by the experimenter. Our protocol allows, in such situations, for the estimation of all the relevant Bargmann invariant phases. 

The usefulness of enhancing generic Hadamard tests---not only cycle tests---by also measuring subsystems that are usually left unmeasured was also recently investigated in Ref.~\cite{faehrmann2025shadowhadamardtestusing}. Although the authors do not focus on the problem of multivariate trace estimation, their analysis shows that the Hadamard test can be powered by local (or global) classical shadows of the quantum system usually left unmeasured to extract additional useful information. 

We have also shown how to estimate third-order invariants of pure states using multiple instances of the destructive swap test. However, we find limited practical relevance for this approach, as the required number of circuit implementations is comparable to performing full state tomography on the two unknown, given qubit states. This represents evidence that variations of the destructive swap test---unlike its non-destructive counterpart, the standard swap test---may not efficiently leverage the trade-off between classical prior information and quantum resources for estimating higher-order invariants. Still, we have described how to implement \emph{destructive} cycle tests, and presented a quantum circuit implementation for estimating third-order invariants.

Future work could explore experimental implementations on existing hardware (e.g., superconducting or photonic platforms) and extend the protocol for the destructive swap test to mixed states of arbitrary dimension. Another interesting direction concerns protocols exploiting controlled causal order of measurements, for example, in a experimentally implementable setup known as the quantum SWITCH~\cite{chiribella2009beyond, Chiribella2013}, which has found various applications in quantum theory~\cite{Chiribella2012, Araujo2014, Felce2020, Zhao2020, Chiribella2021, Wechs2021, Zhu2021, Bavaresco2021, Simonov2022, Simonov2022_Erg, Koudia2022, Liu2023, CalSimCac-23, HectorWood2023, Simonov2023, Liu2024_Deutsch, Rozema2024}. In this setup, quantities like out-of-time-correlators~\cite{Swingle2016} and incompatibility of quantum observables~\cite{gao2023measuring} can be efficiently estimated. These advances would further solidify the role of Bargmann invariants as versatile tools for the certification of quantum resources in a basis-independent way.

\begin{acknowledgments}
RW and EFG acknowledge support from FCT – Fundaç\~{a}o para a Ciência e a Tecnologia (Portugal) via PhD Grant SFRH/BD/151199/2021 and project CEECINST/00062/2018, respectively. This work was also supported by the Digital Horizon Europe project \href{https://cordis.europa.eu/project/id/101070558}{FoQaCiA}, \emph{Foundations of Quantum Computational Advantage}, GA no. 101070558.  RW also acknowledges support from the European Research Council (ERC) under the European Union's Horizon 2020 research and innovation programme (grant agreement No. 856432, HyperQ). This research was funded in whole or in part by the Austrian Science Fund (FWF) 10.55776/PAT4559623. For open access purposes, the author has applied a CC BY public copyright license to any author-accepted manuscript version arising from this submission.
\end{acknowledgments}

\bibliography{references}

\begin{thebibliography}{106}%
\makeatletter
\providecommand \@ifxundefined [1]{%
 \@ifx{#1\undefined}
}%
\providecommand \@ifnum [1]{%
 \ifnum #1\expandafter \@firstoftwo
 \else \expandafter \@secondoftwo
 \fi
}%
\providecommand \@ifx [1]{%
 \ifx #1\expandafter \@firstoftwo
 \else \expandafter \@secondoftwo
 \fi
}%
\providecommand \natexlab [1]{#1}%
\providecommand \enquote  [1]{``#1''}%
\providecommand \bibnamefont  [1]{#1}%
\providecommand \bibfnamefont [1]{#1}%
\providecommand \citenamefont [1]{#1}%
\providecommand \href@noop [0]{\@secondoftwo}%
\providecommand \href [0]{\begingroup \@sanitize@url \@href}%
\providecommand \@href[1]{\@@startlink{#1}\@@href}%
\providecommand \@@href[1]{\endgroup#1\@@endlink}%
\providecommand \@sanitize@url [0]{\catcode `\\12\catcode `\$12\catcode `\&12\catcode `\#12\catcode `\^12\catcode `\_12\catcode `\%12\relax}%
\providecommand \@@startlink[1]{}%
\providecommand \@@endlink[0]{}%
\providecommand \url  [0]{\begingroup\@sanitize@url \@url }%
\providecommand \@url [1]{\endgroup\@href {#1}{\urlprefix }}%
\providecommand \urlprefix  [0]{URL }%
\providecommand \Eprint [0]{\href }%
\providecommand \doibase [0]{https://doi.org/}%
\providecommand \selectlanguage [0]{\@gobble}%
\providecommand \bibinfo  [0]{\@secondoftwo}%
\providecommand \bibfield  [0]{\@secondoftwo}%
\providecommand \translation [1]{[#1]}%
\providecommand \BibitemOpen [0]{}%
\providecommand \bibitemStop [0]{}%
\providecommand \bibitemNoStop [0]{.\EOS\space}%
\providecommand \EOS [0]{\spacefactor3000\relax}%
\providecommand \BibitemShut  [1]{\csname bibitem#1\endcsname}%
\let\auto@bib@innerbib\@empty
\bibitem [{\citenamefont {Bargmann}(1964)}]{bargmann1964note}%
  \BibitemOpen
  \bibfield  {author} {\bibinfo {author} {\bibfnamefont {V.}~\bibnamefont {Bargmann}},\ }\bibfield  {title} {\bibinfo {title} {{Note on Wigner’s Theorem on Symmetry Operations}},\ }\href {https://doi.org/10.1063/1.1704188} {\bibfield  {journal} {\bibinfo  {journal} {J. Math. Phys.}\ }\textbf {\bibinfo {volume} {5}},\ \bibinfo {pages} {862} (\bibinfo {year} {1964})}\BibitemShut {NoStop}%
\bibitem [{\citenamefont {Simon}\ and\ \citenamefont {Mukunda}(1993)}]{simon1993Bargmann}%
  \BibitemOpen
  \bibfield  {author} {\bibinfo {author} {\bibfnamefont {R.}~\bibnamefont {Simon}}\ and\ \bibinfo {author} {\bibfnamefont {N.}~\bibnamefont {Mukunda}},\ }\bibfield  {title} {\bibinfo {title} {{Bargmann invariant and the geometry of the G\"uoy effect}},\ }\href {https://doi.org/10.1103/PhysRevLett.70.880} {\bibfield  {journal} {\bibinfo  {journal} {Phys. Rev. Lett.}\ }\textbf {\bibinfo {volume} {70}},\ \bibinfo {pages} {880} (\bibinfo {year} {1993})}\BibitemShut {NoStop}%
\bibitem [{\citenamefont {Mukunda}\ \emph {et~al.}(2001)\citenamefont {Mukunda}, \citenamefont {Arvind}, \citenamefont {Chaturvedi},\ and\ \citenamefont {Simon}}]{mukunda2001Bargmann}%
  \BibitemOpen
  \bibfield  {author} {\bibinfo {author} {\bibfnamefont {N.}~\bibnamefont {Mukunda}}, \bibinfo {author} {\bibnamefont {Arvind}}, \bibinfo {author} {\bibfnamefont {S.}~\bibnamefont {Chaturvedi}},\ and\ \bibinfo {author} {\bibfnamefont {R.}~\bibnamefont {Simon}},\ }\bibfield  {title} {\bibinfo {title} {Bargmann invariants and off-diagonal geometric phases for multilevel quantum systems: A unitary-group approach},\ }\href {https://doi.org/10.1103/PhysRevA.65.012102} {\bibfield  {journal} {\bibinfo  {journal} {Phys. Rev. A}\ }\textbf {\bibinfo {volume} {65}},\ \bibinfo {pages} {012102} (\bibinfo {year} {2001})}\BibitemShut {NoStop}%
\bibitem [{\citenamefont {Mukunda}\ \emph {et~al.}(2003{\natexlab{a}})\citenamefont {Mukunda}, \citenamefont {Arvind}, \citenamefont {Ercolessi}, \citenamefont {Marmo}, \citenamefont {Morandi},\ and\ \citenamefont {Simon}}]{mukunda2003Bargmann}%
  \BibitemOpen
  \bibfield  {author} {\bibinfo {author} {\bibfnamefont {N.}~\bibnamefont {Mukunda}}, \bibinfo {author} {\bibnamefont {Arvind}}, \bibinfo {author} {\bibfnamefont {E.}~\bibnamefont {Ercolessi}}, \bibinfo {author} {\bibfnamefont {G.}~\bibnamefont {Marmo}}, \bibinfo {author} {\bibfnamefont {G.}~\bibnamefont {Morandi}},\ and\ \bibinfo {author} {\bibfnamefont {R.}~\bibnamefont {Simon}},\ }\bibfield  {title} {\bibinfo {title} {Bargmann invariants, null phase curves, and a theory of the geometric phase},\ }\href {https://doi.org/10.1103/PhysRevA.67.042114} {\bibfield  {journal} {\bibinfo  {journal} {Phys. Rev. A}\ }\textbf {\bibinfo {volume} {67}},\ \bibinfo {pages} {042114} (\bibinfo {year} {2003}{\natexlab{a}})}\BibitemShut {NoStop}%
\bibitem [{\citenamefont {Mukunda}\ \emph {et~al.}(2003{\natexlab{b}})\citenamefont {Mukunda}, \citenamefont {Aravind},\ and\ \citenamefont {Simon}}]{mukunda2003Wigner}%
  \BibitemOpen
  \bibfield  {author} {\bibinfo {author} {\bibfnamefont {N.}~\bibnamefont {Mukunda}}, \bibinfo {author} {\bibfnamefont {P.~K.}\ \bibnamefont {Aravind}},\ and\ \bibinfo {author} {\bibfnamefont {R.}~\bibnamefont {Simon}},\ }\bibfield  {title} {\bibinfo {title} {{Wigner rotations, Bargmann invariants and geometric phases}},\ }\href {https://doi.org/10.1088/0305-4470/36/9/312} {\bibfield  {journal} {\bibinfo  {journal} {J. Phys. A: Math. Gen.}\ }\textbf {\bibinfo {volume} {36}},\ \bibinfo {pages} {2347} (\bibinfo {year} {2003}{\natexlab{b}})}\BibitemShut {NoStop}%
\bibitem [{\citenamefont {Akhilesh}\ \emph {et~al.}(2020)\citenamefont {Akhilesh}, \citenamefont {Arvind}, \citenamefont {Chaturvedi}, \citenamefont {Mallesh},\ and\ \citenamefont {Mukunda}}]{akhilesh2020geometric}%
  \BibitemOpen
  \bibfield  {author} {\bibinfo {author} {\bibfnamefont {K.~S.}\ \bibnamefont {Akhilesh}}, \bibinfo {author} {\bibnamefont {Arvind}}, \bibinfo {author} {\bibfnamefont {S.}~\bibnamefont {Chaturvedi}}, \bibinfo {author} {\bibfnamefont {K.~S.}\ \bibnamefont {Mallesh}},\ and\ \bibinfo {author} {\bibfnamefont {N.}~\bibnamefont {Mukunda}},\ }\bibfield  {title} {\bibinfo {title} {Geometric phases for finite-dimensional systems—the roles of {B}argmann invariants, null phase curves, and the {S}chwinger–{M}ajorana {SU}(2) framework},\ }\href {https://doi.org/10.1063/1.5124865} {\bibfield  {journal} {\bibinfo  {journal} {J. Math. Phys.}\ }\textbf {\bibinfo {volume} {61}},\ \bibinfo {pages} {072103} (\bibinfo {year} {2020})}\BibitemShut {NoStop}%
\bibitem [{\citenamefont {Fano}(1957)}]{fano1957description}%
  \BibitemOpen
  \bibfield  {author} {\bibinfo {author} {\bibfnamefont {U.}~\bibnamefont {Fano}},\ }\bibfield  {title} {\bibinfo {title} {{Description of States in Quantum Mechanics by Density Matrix and Operator Techniques}},\ }\href {https://doi.org/10.1103/RevModPhys.29.74} {\bibfield  {journal} {\bibinfo  {journal} {Rev. Mod. Phys.}\ }\textbf {\bibinfo {volume} {29}},\ \bibinfo {pages} {74} (\bibinfo {year} {1957})}\BibitemShut {NoStop}%
\bibitem [{Note1()}]{Note1}%
  \BibitemOpen
  \bibinfo {note} {If we write $\Delta _n(\protect \pmb \varrho ) = \Delta e^{i\phi _\Delta }$, with $\Delta \in \protect \mathbb {R}_{\geq 0}$, the value $\phi _\Delta $ is called the \protect \emph {phase} of the invariant.}\BibitemShut {Stop}%
\bibitem [{\citenamefont {Shchesnovich}(2015)}]{shchesnovich2015partial}%
  \BibitemOpen
  \bibfield  {author} {\bibinfo {author} {\bibfnamefont {V.~S.}\ \bibnamefont {Shchesnovich}},\ }\bibfield  {title} {\bibinfo {title} {Partial indistinguishability theory for multiphoton experiments in multiport devices},\ }\href {https://doi.org/10.1103/PhysRevA.91.013844} {\bibfield  {journal} {\bibinfo  {journal} {Phys. Rev. A}\ }\textbf {\bibinfo {volume} {91}},\ \bibinfo {pages} {013844} (\bibinfo {year} {2015})}\BibitemShut {NoStop}%
\bibitem [{\citenamefont {Shchesnovich}\ and\ \citenamefont {Bezerra}(2018)}]{shchesnovich2018collective}%
  \BibitemOpen
  \bibfield  {author} {\bibinfo {author} {\bibfnamefont {V.~S.}\ \bibnamefont {Shchesnovich}}\ and\ \bibinfo {author} {\bibfnamefont {M.~E.~O.}\ \bibnamefont {Bezerra}},\ }\bibfield  {title} {\bibinfo {title} {Collective phases of identical particles interfering on linear multiports},\ }\href {https://doi.org/10.1103/PhysRevA.98.033805} {\bibfield  {journal} {\bibinfo  {journal} {Phys. Rev. A}\ }\textbf {\bibinfo {volume} {98}},\ \bibinfo {pages} {033805} (\bibinfo {year} {2018})}\BibitemShut {NoStop}%
\bibitem [{\citenamefont {Bong}\ \emph {et~al.}(2018)\citenamefont {Bong}, \citenamefont {Tischler}, \citenamefont {Patel}, \citenamefont {Wollmann}, \citenamefont {Pryde},\ and\ \citenamefont {Hall}}]{bong2018strong}%
  \BibitemOpen
  \bibfield  {author} {\bibinfo {author} {\bibfnamefont {K.-W.}\ \bibnamefont {Bong}}, \bibinfo {author} {\bibfnamefont {N.}~\bibnamefont {Tischler}}, \bibinfo {author} {\bibfnamefont {R.~B.}\ \bibnamefont {Patel}}, \bibinfo {author} {\bibfnamefont {S.}~\bibnamefont {Wollmann}}, \bibinfo {author} {\bibfnamefont {G.~J.}\ \bibnamefont {Pryde}},\ and\ \bibinfo {author} {\bibfnamefont {M.~J.~W.}\ \bibnamefont {Hall}},\ }\bibfield  {title} {\bibinfo {title} {{Strong Unitary and Overlap Uncertainty Relations: Theory and Experiment}},\ }\href {https://doi.org/10.1103/PhysRevLett.120.230402} {\bibfield  {journal} {\bibinfo  {journal} {Phys. Rev. Lett.}\ }\textbf {\bibinfo {volume} {120}},\ \bibinfo {pages} {230402} (\bibinfo {year} {2018})}\BibitemShut {NoStop}%
\bibitem [{\citenamefont {Jones}\ \emph {et~al.}(2020)\citenamefont {Jones}, \citenamefont {Menssen}, \citenamefont {Chrzanowski}, \citenamefont {Wolterink}, \citenamefont {Shchesnovich},\ and\ \citenamefont {Walmsley}}]{jones2020multiparticle}%
  \BibitemOpen
  \bibfield  {author} {\bibinfo {author} {\bibfnamefont {A.~E.}\ \bibnamefont {Jones}}, \bibinfo {author} {\bibfnamefont {A.~J.}\ \bibnamefont {Menssen}}, \bibinfo {author} {\bibfnamefont {H.~M.}\ \bibnamefont {Chrzanowski}}, \bibinfo {author} {\bibfnamefont {T.~A.~W.}\ \bibnamefont {Wolterink}}, \bibinfo {author} {\bibfnamefont {V.~S.}\ \bibnamefont {Shchesnovich}},\ and\ \bibinfo {author} {\bibfnamefont {I.~A.}\ \bibnamefont {Walmsley}},\ }\bibfield  {title} {\bibinfo {title} {{Multiparticle Interference of Pairwise Distinguishable Photons}},\ }\href {https://doi.org/10.1103/PhysRevLett.125.123603} {\bibfield  {journal} {\bibinfo  {journal} {Phys. Rev. Lett.}\ }\textbf {\bibinfo {volume} {125}},\ \bibinfo {pages} {123603} (\bibinfo {year} {2020})}\BibitemShut {NoStop}%
\bibitem [{\citenamefont {Pont}\ \emph {et~al.}(2022)\citenamefont {Pont}, \citenamefont {Albiero}, \citenamefont {Thomas}, \citenamefont {Spagnolo}, \citenamefont {Ceccarelli}, \citenamefont {Corrielli}, \citenamefont {Brieussel}, \citenamefont {Somaschi}, \citenamefont {Huet}, \citenamefont {Harouri}, \citenamefont {Lema\^{\i}tre}, \citenamefont {Sagnes}, \citenamefont {Belabas}, \citenamefont {Sciarrino}, \citenamefont {Osellame}, \citenamefont {Senellart},\ and\ \citenamefont {Crespi}}]{pont2022quantifying}%
  \BibitemOpen
  \bibfield  {author} {\bibinfo {author} {\bibfnamefont {M.}~\bibnamefont {Pont}}, \bibinfo {author} {\bibfnamefont {R.}~\bibnamefont {Albiero}}, \bibinfo {author} {\bibfnamefont {S.~E.}\ \bibnamefont {Thomas}}, \bibinfo {author} {\bibfnamefont {N.}~\bibnamefont {Spagnolo}}, \bibinfo {author} {\bibfnamefont {F.}~\bibnamefont {Ceccarelli}}, \bibinfo {author} {\bibfnamefont {G.}~\bibnamefont {Corrielli}}, \bibinfo {author} {\bibfnamefont {A.}~\bibnamefont {Brieussel}}, \bibinfo {author} {\bibfnamefont {N.}~\bibnamefont {Somaschi}}, \bibinfo {author} {\bibfnamefont {H.}~\bibnamefont {Huet}}, \bibinfo {author} {\bibfnamefont {A.}~\bibnamefont {Harouri}}, \bibinfo {author} {\bibfnamefont {A.}~\bibnamefont {Lema\^{\i}tre}}, \bibinfo {author} {\bibfnamefont {I.}~\bibnamefont {Sagnes}}, \bibinfo {author} {\bibfnamefont {N.}~\bibnamefont {Belabas}}, \bibinfo {author} {\bibfnamefont {F.}~\bibnamefont {Sciarrino}}, \bibinfo {author} {\bibfnamefont {R.}~\bibnamefont {Osellame}}, \bibinfo {author} {\bibfnamefont
  {P.}~\bibnamefont {Senellart}},\ and\ \bibinfo {author} {\bibfnamefont {A.}~\bibnamefont {Crespi}},\ }\bibfield  {title} {\bibinfo {title} {{Quantifying $n$-Photon Indistinguishability with a Cyclic Integrated Interferometer}},\ }\href {https://doi.org/10.1103/PhysRevX.12.031033} {\bibfield  {journal} {\bibinfo  {journal} {Phys. Rev. X}\ }\textbf {\bibinfo {volume} {12}},\ \bibinfo {pages} {031033} (\bibinfo {year} {2022})}\BibitemShut {NoStop}%
\bibitem [{\citenamefont {Jones}\ \emph {et~al.}(2023)\citenamefont {Jones}, \citenamefont {Kumar}, \citenamefont {D'Aurelio}, \citenamefont {Bayerbach}, \citenamefont {Menssen},\ and\ \citenamefont {Barz}}]{jones2023distinguishability}%
  \BibitemOpen
  \bibfield  {author} {\bibinfo {author} {\bibfnamefont {A.~E.}\ \bibnamefont {Jones}}, \bibinfo {author} {\bibfnamefont {S.}~\bibnamefont {Kumar}}, \bibinfo {author} {\bibfnamefont {S.}~\bibnamefont {D'Aurelio}}, \bibinfo {author} {\bibfnamefont {M.}~\bibnamefont {Bayerbach}}, \bibinfo {author} {\bibfnamefont {A.~J.}\ \bibnamefont {Menssen}},\ and\ \bibinfo {author} {\bibfnamefont {S.}~\bibnamefont {Barz}},\ }\bibfield  {title} {\bibinfo {title} {Distinguishability and mixedness in quantum interference},\ }\href {https://doi.org/10.1103/PhysRevA.108.053701} {\bibfield  {journal} {\bibinfo  {journal} {Phys. Rev. A}\ }\textbf {\bibinfo {volume} {108}},\ \bibinfo {pages} {053701} (\bibinfo {year} {2023})}\BibitemShut {NoStop}%
\bibitem [{\citenamefont {Rodari}\ \emph {et~al.}(2024{\natexlab{a}})\citenamefont {Rodari}, \citenamefont {Fernandes}, \citenamefont {Caruccio}, \citenamefont {Suprano}, \citenamefont {Hoch}, \citenamefont {Giordani}, \citenamefont {Carvacho}, \citenamefont {Albiero}, \citenamefont {Giano}, \citenamefont {Corrielli}, \citenamefont {Ceccarelli}, \citenamefont {Osellame}, \citenamefont {Brod}, \citenamefont {Novo}, \citenamefont {Spagnolo}, \citenamefont {Galvão},\ and\ \citenamefont {Sciarrino}}]{rodari2024experimentalobservationcounterintuitivefeatures}%
  \BibitemOpen
  \bibfield  {author} {\bibinfo {author} {\bibfnamefont {G.}~\bibnamefont {Rodari}}, \bibinfo {author} {\bibfnamefont {C.}~\bibnamefont {Fernandes}}, \bibinfo {author} {\bibfnamefont {E.}~\bibnamefont {Caruccio}}, \bibinfo {author} {\bibfnamefont {A.}~\bibnamefont {Suprano}}, \bibinfo {author} {\bibfnamefont {F.}~\bibnamefont {Hoch}}, \bibinfo {author} {\bibfnamefont {T.}~\bibnamefont {Giordani}}, \bibinfo {author} {\bibfnamefont {G.}~\bibnamefont {Carvacho}}, \bibinfo {author} {\bibfnamefont {R.}~\bibnamefont {Albiero}}, \bibinfo {author} {\bibfnamefont {N.~D.}\ \bibnamefont {Giano}}, \bibinfo {author} {\bibfnamefont {G.}~\bibnamefont {Corrielli}}, \bibinfo {author} {\bibfnamefont {F.}~\bibnamefont {Ceccarelli}}, \bibinfo {author} {\bibfnamefont {R.}~\bibnamefont {Osellame}}, \bibinfo {author} {\bibfnamefont {D.~J.}\ \bibnamefont {Brod}}, \bibinfo {author} {\bibfnamefont {L.}~\bibnamefont {Novo}}, \bibinfo {author} {\bibfnamefont {N.}~\bibnamefont {Spagnolo}}, \bibinfo {author} {\bibfnamefont {E.~F.}\
  \bibnamefont {Galvão}},\ and\ \bibinfo {author} {\bibfnamefont {F.}~\bibnamefont {Sciarrino}},\ }\href {https://doi.org/https://doi.org/10.48550/arXiv.2410.15883} {\bibinfo {title} {Experimental observation of counter-intuitive features of photonic bunching}} (\bibinfo {year} {2024}{\natexlab{a}}),\ \Eprint {https://arxiv.org/abs/2410.15883} {arXiv:2410.15883 [quant-ph]} \BibitemShut {NoStop}%
\bibitem [{\citenamefont {Rodari}\ \emph {et~al.}(2024{\natexlab{b}})\citenamefont {Rodari}, \citenamefont {Novo}, \citenamefont {Albiero}, \citenamefont {Suprano}, \citenamefont {Tavares}, \citenamefont {Caruccio}, \citenamefont {Hoch}, \citenamefont {Giordani}, \citenamefont {Carvacho}, \citenamefont {Gardina}, \citenamefont {Giano}, \citenamefont {Giorgio}, \citenamefont {Corrielli}, \citenamefont {Ceccarelli}, \citenamefont {Osellame}, \citenamefont {Spagnolo}, \citenamefont {Galvão},\ and\ \citenamefont {Sciarrino}}]{rodari2024semideviceindependentcharacterizationmultiphoton}%
  \BibitemOpen
  \bibfield  {author} {\bibinfo {author} {\bibfnamefont {G.}~\bibnamefont {Rodari}}, \bibinfo {author} {\bibfnamefont {L.}~\bibnamefont {Novo}}, \bibinfo {author} {\bibfnamefont {R.}~\bibnamefont {Albiero}}, \bibinfo {author} {\bibfnamefont {A.}~\bibnamefont {Suprano}}, \bibinfo {author} {\bibfnamefont {C.~T.}\ \bibnamefont {Tavares}}, \bibinfo {author} {\bibfnamefont {E.}~\bibnamefont {Caruccio}}, \bibinfo {author} {\bibfnamefont {F.}~\bibnamefont {Hoch}}, \bibinfo {author} {\bibfnamefont {T.}~\bibnamefont {Giordani}}, \bibinfo {author} {\bibfnamefont {G.}~\bibnamefont {Carvacho}}, \bibinfo {author} {\bibfnamefont {M.}~\bibnamefont {Gardina}}, \bibinfo {author} {\bibfnamefont {N.~D.}\ \bibnamefont {Giano}}, \bibinfo {author} {\bibfnamefont {S.~D.}\ \bibnamefont {Giorgio}}, \bibinfo {author} {\bibfnamefont {G.}~\bibnamefont {Corrielli}}, \bibinfo {author} {\bibfnamefont {F.}~\bibnamefont {Ceccarelli}}, \bibinfo {author} {\bibfnamefont {R.}~\bibnamefont {Osellame}}, \bibinfo {author} {\bibfnamefont
  {N.}~\bibnamefont {Spagnolo}}, \bibinfo {author} {\bibfnamefont {E.~F.}\ \bibnamefont {Galvão}},\ and\ \bibinfo {author} {\bibfnamefont {F.}~\bibnamefont {Sciarrino}},\ }\href {https://doi.org/https://doi.org/10.48550/arXiv.2404.18636} {\bibinfo {title} {Semi-device independent characterization of multiphoton indistinguishability}} (\bibinfo {year} {2024}{\natexlab{b}}),\ \Eprint {https://arxiv.org/abs/2404.18636} {arXiv:2404.18636 [quant-ph]} \BibitemShut {NoStop}%
\bibitem [{\citenamefont {Oszmaniec}\ \emph {et~al.}(2024)\citenamefont {Oszmaniec}, \citenamefont {Brod},\ and\ \citenamefont {Galvão}}]{oszmaniec2024measuring}%
  \BibitemOpen
  \bibfield  {author} {\bibinfo {author} {\bibfnamefont {M.}~\bibnamefont {Oszmaniec}}, \bibinfo {author} {\bibfnamefont {D.~J.}\ \bibnamefont {Brod}},\ and\ \bibinfo {author} {\bibfnamefont {E.~F.}\ \bibnamefont {Galvão}},\ }\bibfield  {title} {\bibinfo {title} {Measuring relational information between quantum states, and applications},\ }\href {https://doi.org/10.1088/1367-2630/ad1a27} {\bibfield  {journal} {\bibinfo  {journal} {New J. Phys.}\ }\textbf {\bibinfo {volume} {26}},\ \bibinfo {pages} {013053} (\bibinfo {year} {2024})}\BibitemShut {NoStop}%
\bibitem [{\citenamefont {Galv\~ao}\ and\ \citenamefont {Brod}(2020)}]{galvao2020quantumandclassical}%
  \BibitemOpen
  \bibfield  {author} {\bibinfo {author} {\bibfnamefont {E.~F.}\ \bibnamefont {Galv\~ao}}\ and\ \bibinfo {author} {\bibfnamefont {D.~J.}\ \bibnamefont {Brod}},\ }\bibfield  {title} {\bibinfo {title} {Quantum and classical bounds for two-state overlaps},\ }\href {https://doi.org/10.1103/PhysRevA.101.062110} {\bibfield  {journal} {\bibinfo  {journal} {Phys. Rev. A}\ }\textbf {\bibinfo {volume} {101}},\ \bibinfo {pages} {062110} (\bibinfo {year} {2020})}\BibitemShut {NoStop}%
\bibitem [{\citenamefont {Giordani}\ \emph {et~al.}(2021)\citenamefont {Giordani}, \citenamefont {Esposito}, \citenamefont {Hoch}, \citenamefont {Carvacho}, \citenamefont {Brod}, \citenamefont {Galv\~ao}, \citenamefont {Spagnolo},\ and\ \citenamefont {Sciarrino}}]{giordani2021witnessing}%
  \BibitemOpen
  \bibfield  {author} {\bibinfo {author} {\bibfnamefont {T.}~\bibnamefont {Giordani}}, \bibinfo {author} {\bibfnamefont {C.}~\bibnamefont {Esposito}}, \bibinfo {author} {\bibfnamefont {F.}~\bibnamefont {Hoch}}, \bibinfo {author} {\bibfnamefont {G.}~\bibnamefont {Carvacho}}, \bibinfo {author} {\bibfnamefont {D.~J.}\ \bibnamefont {Brod}}, \bibinfo {author} {\bibfnamefont {E.~F.}\ \bibnamefont {Galv\~ao}}, \bibinfo {author} {\bibfnamefont {N.}~\bibnamefont {Spagnolo}},\ and\ \bibinfo {author} {\bibfnamefont {F.}~\bibnamefont {Sciarrino}},\ }\bibfield  {title} {\bibinfo {title} {Witnesses of coherence and dimension from multiphoton indistinguishability tests},\ }\href {https://doi.org/10.1103/PhysRevResearch.3.023031} {\bibfield  {journal} {\bibinfo  {journal} {Phys. Rev. Research}\ }\textbf {\bibinfo {volume} {3}},\ \bibinfo {pages} {023031} (\bibinfo {year} {2021})}\BibitemShut {NoStop}%
\bibitem [{\citenamefont {Wagner}\ \emph {et~al.}(2024{\natexlab{a}})\citenamefont {Wagner}, \citenamefont {Barbosa},\ and\ \citenamefont {Galv\~ao}}]{wagner2024inequalities}%
  \BibitemOpen
  \bibfield  {author} {\bibinfo {author} {\bibfnamefont {R.}~\bibnamefont {Wagner}}, \bibinfo {author} {\bibfnamefont {R.~S.}\ \bibnamefont {Barbosa}},\ and\ \bibinfo {author} {\bibfnamefont {E.~F.}\ \bibnamefont {Galv\~ao}},\ }\bibfield  {title} {\bibinfo {title} {Inequalities witnessing coherence, nonlocality, and contextuality},\ }\href {https://doi.org/10.1103/PhysRevA.109.032220} {\bibfield  {journal} {\bibinfo  {journal} {Phys. Rev. A}\ }\textbf {\bibinfo {volume} {109}},\ \bibinfo {pages} {032220} (\bibinfo {year} {2024}{\natexlab{a}})}\BibitemShut {NoStop}%
\bibitem [{\citenamefont {Wagner}\ \emph {et~al.}(2024{\natexlab{b}})\citenamefont {Wagner}, \citenamefont {Camillini},\ and\ \citenamefont {Galv{\~{a}}o}}]{wagner2024coherence}%
  \BibitemOpen
  \bibfield  {author} {\bibinfo {author} {\bibfnamefont {R.}~\bibnamefont {Wagner}}, \bibinfo {author} {\bibfnamefont {A.}~\bibnamefont {Camillini}},\ and\ \bibinfo {author} {\bibfnamefont {E.~F.}\ \bibnamefont {Galv{\~{a}}o}},\ }\bibfield  {title} {\bibinfo {title} {Coherence and contextuality in a {M}ach-{Z}ehnder interferometer},\ }\href {https://doi.org/10.22331/q-2024-02-05-1240} {\bibfield  {journal} {\bibinfo  {journal} {{Quantum}}\ }\textbf {\bibinfo {volume} {8}},\ \bibinfo {pages} {1240} (\bibinfo {year} {2024}{\natexlab{b}})}\BibitemShut {NoStop}%
\bibitem [{\citenamefont {Wagner}\ \emph {et~al.}(2024{\natexlab{c}})\citenamefont {Wagner}, \citenamefont {Peres}, \citenamefont {Cruzeiro},\ and\ \citenamefont {Galvão}}]{wagner2024certifying}%
  \BibitemOpen
  \bibfield  {author} {\bibinfo {author} {\bibfnamefont {R.}~\bibnamefont {Wagner}}, \bibinfo {author} {\bibfnamefont {F.~C.~R.}\ \bibnamefont {Peres}}, \bibinfo {author} {\bibfnamefont {E.~Z.}\ \bibnamefont {Cruzeiro}},\ and\ \bibinfo {author} {\bibfnamefont {E.~F.}\ \bibnamefont {Galvão}},\ }\href {https://doi.org/https://doi.org/10.48550/arXiv.2404.16107} {\bibinfo {title} {Certifying nonstabilizerness in quantum processors}} (\bibinfo {year} {2024}{\natexlab{c}}),\ \Eprint {https://arxiv.org/abs/2404.16107} {arXiv:2404.16107 [quant-ph]} \BibitemShut {NoStop}%
\bibitem [{\citenamefont {Haug}\ and\ \citenamefont {Kim}(2023)}]{haug2023scalable}%
  \BibitemOpen
  \bibfield  {author} {\bibinfo {author} {\bibfnamefont {T.}~\bibnamefont {Haug}}\ and\ \bibinfo {author} {\bibfnamefont {M.}~\bibnamefont {Kim}},\ }\bibfield  {title} {\bibinfo {title} {{Scalable Measures of Magic Resource for Quantum Computers}},\ }\href {https://doi.org/10.1103/PRXQuantum.4.010301} {\bibfield  {journal} {\bibinfo  {journal} {PRX Quantum}\ }\textbf {\bibinfo {volume} {4}},\ \bibinfo {pages} {010301} (\bibinfo {year} {2023})}\BibitemShut {NoStop}%
\bibitem [{\citenamefont {Giordani}\ \emph {et~al.}(2023)\citenamefont {Giordani}, \citenamefont {Wagner}, \citenamefont {Esposito}, \citenamefont {Camillini}, \citenamefont {Hoch}, \citenamefont {Carvacho}, \citenamefont {Pentangelo}, \citenamefont {Ceccarelli}, \citenamefont {Piacentini}, \citenamefont {Crespi}, \citenamefont {Spagnolo}, \citenamefont {Osellame}, \citenamefont {Galvão},\ and\ \citenamefont {Sciarrino}}]{giordani2023experimentalcertification}%
  \BibitemOpen
  \bibfield  {author} {\bibinfo {author} {\bibfnamefont {T.}~\bibnamefont {Giordani}}, \bibinfo {author} {\bibfnamefont {R.}~\bibnamefont {Wagner}}, \bibinfo {author} {\bibfnamefont {C.}~\bibnamefont {Esposito}}, \bibinfo {author} {\bibfnamefont {A.}~\bibnamefont {Camillini}}, \bibinfo {author} {\bibfnamefont {F.}~\bibnamefont {Hoch}}, \bibinfo {author} {\bibfnamefont {G.}~\bibnamefont {Carvacho}}, \bibinfo {author} {\bibfnamefont {C.}~\bibnamefont {Pentangelo}}, \bibinfo {author} {\bibfnamefont {F.}~\bibnamefont {Ceccarelli}}, \bibinfo {author} {\bibfnamefont {S.}~\bibnamefont {Piacentini}}, \bibinfo {author} {\bibfnamefont {A.}~\bibnamefont {Crespi}}, \bibinfo {author} {\bibfnamefont {N.}~\bibnamefont {Spagnolo}}, \bibinfo {author} {\bibfnamefont {R.}~\bibnamefont {Osellame}}, \bibinfo {author} {\bibfnamefont {E.~F.}\ \bibnamefont {Galvão}},\ and\ \bibinfo {author} {\bibfnamefont {F.}~\bibnamefont {Sciarrino}},\ }\bibfield  {title} {\bibinfo {title} {Experimental certification of contextuality, coherence,
  and dimension in a programmable universal photonic processor},\ }\href {https://doi.org/10.1126/sciadv.adj4249} {\bibfield  {journal} {\bibinfo  {journal} {Science Advances}\ }\textbf {\bibinfo {volume} {9}},\ \bibinfo {pages} {eadj4249} (\bibinfo {year} {2023})}\BibitemShut {NoStop}%
\bibitem [{\citenamefont {Brod}\ \emph {et~al.}(2019)\citenamefont {Brod}, \citenamefont {Galv\~ao}, \citenamefont {Viggianiello}, \citenamefont {Flamini}, \citenamefont {Spagnolo},\ and\ \citenamefont {Sciarrino}}]{brod2019witnessing}%
  \BibitemOpen
  \bibfield  {author} {\bibinfo {author} {\bibfnamefont {D.~J.}\ \bibnamefont {Brod}}, \bibinfo {author} {\bibfnamefont {E.~F.}\ \bibnamefont {Galv\~ao}}, \bibinfo {author} {\bibfnamefont {N.}~\bibnamefont {Viggianiello}}, \bibinfo {author} {\bibfnamefont {F.}~\bibnamefont {Flamini}}, \bibinfo {author} {\bibfnamefont {N.}~\bibnamefont {Spagnolo}},\ and\ \bibinfo {author} {\bibfnamefont {F.}~\bibnamefont {Sciarrino}},\ }\bibfield  {title} {\bibinfo {title} {{Witnessing Genuine Multiphoton Indistinguishability}},\ }\href {https://doi.org/10.1103/PhysRevLett.122.063602} {\bibfield  {journal} {\bibinfo  {journal} {Phys. Rev. Lett.}\ }\textbf {\bibinfo {volume} {122}},\ \bibinfo {pages} {063602} (\bibinfo {year} {2019})}\BibitemShut {NoStop}%
\bibitem [{\citenamefont {Giordani}\ \emph {et~al.}(2020)\citenamefont {Giordani}, \citenamefont {Brod}, \citenamefont {Esposito}, \citenamefont {Viggianiello}, \citenamefont {Romano}, \citenamefont {Flamini}, \citenamefont {Carvacho}, \citenamefont {Spagnolo}, \citenamefont {Galvão},\ and\ \citenamefont {Sciarrino}}]{giordani2020experimentalquantification}%
  \BibitemOpen
  \bibfield  {author} {\bibinfo {author} {\bibfnamefont {T.}~\bibnamefont {Giordani}}, \bibinfo {author} {\bibfnamefont {D.~J.}\ \bibnamefont {Brod}}, \bibinfo {author} {\bibfnamefont {C.}~\bibnamefont {Esposito}}, \bibinfo {author} {\bibfnamefont {N.}~\bibnamefont {Viggianiello}}, \bibinfo {author} {\bibfnamefont {M.}~\bibnamefont {Romano}}, \bibinfo {author} {\bibfnamefont {F.}~\bibnamefont {Flamini}}, \bibinfo {author} {\bibfnamefont {G.}~\bibnamefont {Carvacho}}, \bibinfo {author} {\bibfnamefont {N.}~\bibnamefont {Spagnolo}}, \bibinfo {author} {\bibfnamefont {E.~F.}\ \bibnamefont {Galvão}},\ and\ \bibinfo {author} {\bibfnamefont {F.}~\bibnamefont {Sciarrino}},\ }\bibfield  {title} {\bibinfo {title} {Experimental quantification of four-photon indistinguishability},\ }\href {https://doi.org/10.1088/1367-2630/ab7a30} {\bibfield  {journal} {\bibinfo  {journal} {New J. Phys.}\ }\textbf {\bibinfo {volume} {22}},\ \bibinfo {pages} {043001} (\bibinfo {year} {2020})}\BibitemShut {NoStop}%
\bibitem [{\citenamefont {Miklin}\ and\ \citenamefont {Oszmaniec}(2021)}]{miklin2021universalscheme}%
  \BibitemOpen
  \bibfield  {author} {\bibinfo {author} {\bibfnamefont {N.}~\bibnamefont {Miklin}}\ and\ \bibinfo {author} {\bibfnamefont {M.}~\bibnamefont {Oszmaniec}},\ }\bibfield  {title} {\bibinfo {title} {A universal scheme for robust self-testing in the prepare-and-measure scenario},\ }\href {https://doi.org/10.22331/q-2021-04-06-424} {\bibfield  {journal} {\bibinfo  {journal} {{Quantum}}\ }\textbf {\bibinfo {volume} {5}},\ \bibinfo {pages} {424} (\bibinfo {year} {2021})}\BibitemShut {NoStop}%
\bibitem [{\citenamefont {Hinsche}\ \emph {et~al.}(2024)\citenamefont {Hinsche}, \citenamefont {Ioannou}, \citenamefont {Jerbi}, \citenamefont {Leone}, \citenamefont {Eisert},\ and\ \citenamefont {Carrasco}}]{hinsche2024efficientdistributedinnerproduct}%
  \BibitemOpen
  \bibfield  {author} {\bibinfo {author} {\bibfnamefont {M.}~\bibnamefont {Hinsche}}, \bibinfo {author} {\bibfnamefont {M.}~\bibnamefont {Ioannou}}, \bibinfo {author} {\bibfnamefont {S.}~\bibnamefont {Jerbi}}, \bibinfo {author} {\bibfnamefont {L.}~\bibnamefont {Leone}}, \bibinfo {author} {\bibfnamefont {J.}~\bibnamefont {Eisert}},\ and\ \bibinfo {author} {\bibfnamefont {J.}~\bibnamefont {Carrasco}},\ }\href {https://doi.org/https://doi.org/10.48550/arXiv.2405.06544} {\bibinfo {title} {Efficient distributed inner product estimation via pauli sampling}} (\bibinfo {year} {2024}),\ \Eprint {https://arxiv.org/abs/2405.06544} {arXiv:2405.06544 [quant-ph]} \BibitemShut {NoStop}%
\bibitem [{\citenamefont {Bandyopadhyay}\ \emph {et~al.}(2023)\citenamefont {Bandyopadhyay}, \citenamefont {Rubin}, \citenamefont {Radulaski},\ and\ \citenamefont {Wilde}}]{bandyopadhyay2023efficient}%
  \BibitemOpen
  \bibfield  {author} {\bibinfo {author} {\bibfnamefont {R.}~\bibnamefont {Bandyopadhyay}}, \bibinfo {author} {\bibfnamefont {A.~H.}\ \bibnamefont {Rubin}}, \bibinfo {author} {\bibfnamefont {M.}~\bibnamefont {Radulaski}},\ and\ \bibinfo {author} {\bibfnamefont {M.~M.}\ \bibnamefont {Wilde}},\ }\bibfield  {title} {\bibinfo {title} {{Efficient Quantum Algorithms for Testing Symmetries of Open Quantum Systems}},\ }\href {https://doi.org/10.1142/s1230161223500178} {\bibfield  {journal} {\bibinfo  {journal} {Open Systems and Information Dynamics}\ }\textbf {\bibinfo {volume} {30}},\ \bibinfo {pages} {2350017} (\bibinfo {year} {2023})}\BibitemShut {NoStop}%
\bibitem [{\citenamefont {Huang}\ \emph {et~al.}(2024)\citenamefont {Huang}, \citenamefont {Preskill},\ and\ \citenamefont {Soleimanifar}}]{huang2024certifyingquantumstatessinglequbit}%
  \BibitemOpen
  \bibfield  {author} {\bibinfo {author} {\bibfnamefont {H.-Y.}\ \bibnamefont {Huang}}, \bibinfo {author} {\bibfnamefont {J.}~\bibnamefont {Preskill}},\ and\ \bibinfo {author} {\bibfnamefont {M.}~\bibnamefont {Soleimanifar}},\ }\href {https://doi.org/https://doi.org/10.48550/arXiv.2404.07281} {\bibinfo {title} {Certifying almost all quantum states with few single-qubit measurements}} (\bibinfo {year} {2024}),\ \Eprint {https://arxiv.org/abs/2404.07281} {arXiv:2404.07281 [quant-ph]} \BibitemShut {NoStop}%
\bibitem [{\citenamefont {Tamir}\ and\ \citenamefont {Cohen}(2013)}]{tamir2013introduction}%
  \BibitemOpen
  \bibfield  {author} {\bibinfo {author} {\bibfnamefont {B.}~\bibnamefont {Tamir}}\ and\ \bibinfo {author} {\bibfnamefont {E.}~\bibnamefont {Cohen}},\ }\bibfield  {title} {\bibinfo {title} {{Introduction to Weak Measurements and Weak Values}},\ }\href {https://doi.org/10.12743/quanta.v2i1.14} {\bibfield  {journal} {\bibinfo  {journal} {Quanta}\ }\textbf {\bibinfo {volume} {2}},\ \bibinfo {pages} {7} (\bibinfo {year} {2013})}\BibitemShut {NoStop}%
\bibitem [{\citenamefont {Dressel}\ \emph {et~al.}(2014)\citenamefont {Dressel}, \citenamefont {Malik}, \citenamefont {Miatto}, \citenamefont {Jordan},\ and\ \citenamefont {Boyd}}]{dressel2014colloquium}%
  \BibitemOpen
  \bibfield  {author} {\bibinfo {author} {\bibfnamefont {J.}~\bibnamefont {Dressel}}, \bibinfo {author} {\bibfnamefont {M.}~\bibnamefont {Malik}}, \bibinfo {author} {\bibfnamefont {F.~M.}\ \bibnamefont {Miatto}}, \bibinfo {author} {\bibfnamefont {A.~N.}\ \bibnamefont {Jordan}},\ and\ \bibinfo {author} {\bibfnamefont {R.~W.}\ \bibnamefont {Boyd}},\ }\bibfield  {title} {\bibinfo {title} {Colloquium: Understanding quantum weak values: Basics and applications},\ }\href {https://doi.org/10.1103/RevModPhys.86.307} {\bibfield  {journal} {\bibinfo  {journal} {Rev. Mod. Phys.}\ }\textbf {\bibinfo {volume} {86}},\ \bibinfo {pages} {307} (\bibinfo {year} {2014})}\BibitemShut {NoStop}%
\bibitem [{\citenamefont {Wagner}\ and\ \citenamefont {Galv\~ao}(2023)}]{wagner2023simple}%
  \BibitemOpen
  \bibfield  {author} {\bibinfo {author} {\bibfnamefont {R.}~\bibnamefont {Wagner}}\ and\ \bibinfo {author} {\bibfnamefont {E.~F.}\ \bibnamefont {Galv\~ao}},\ }\bibfield  {title} {\bibinfo {title} {Simple proof that anomalous weak values require coherence},\ }\href {https://doi.org/10.1103/PhysRevA.108.L040202} {\bibfield  {journal} {\bibinfo  {journal} {Phys. Rev. A}\ }\textbf {\bibinfo {volume} {108}},\ \bibinfo {pages} {L040202} (\bibinfo {year} {2023})}\BibitemShut {NoStop}%
\bibitem [{\citenamefont {Lostaglio}\ \emph {et~al.}(2023)\citenamefont {Lostaglio}, \citenamefont {Belenchia}, \citenamefont {Levy}, \citenamefont {Hern{\'{a}}ndez-G{\'{o}}mez}, \citenamefont {Fabbri},\ and\ \citenamefont {Gherardini}}]{lostaglio2023kirkwooddirac}%
  \BibitemOpen
  \bibfield  {author} {\bibinfo {author} {\bibfnamefont {M.}~\bibnamefont {Lostaglio}}, \bibinfo {author} {\bibfnamefont {A.}~\bibnamefont {Belenchia}}, \bibinfo {author} {\bibfnamefont {A.}~\bibnamefont {Levy}}, \bibinfo {author} {\bibfnamefont {S.}~\bibnamefont {Hern{\'{a}}ndez-G{\'{o}}mez}}, \bibinfo {author} {\bibfnamefont {N.}~\bibnamefont {Fabbri}},\ and\ \bibinfo {author} {\bibfnamefont {S.}~\bibnamefont {Gherardini}},\ }\bibfield  {title} {\bibinfo {title} {Kirkwood-{D}irac quasiprobability approach to the statistics of incompatible observables},\ }\href {https://doi.org/10.22331/q-2023-10-09-1128} {\bibfield  {journal} {\bibinfo  {journal} {{Quantum}}\ }\textbf {\bibinfo {volume} {7}},\ \bibinfo {pages} {1128} (\bibinfo {year} {2023})}\BibitemShut {NoStop}%
\bibitem [{\citenamefont {Wagner}\ \emph {et~al.}(2024{\natexlab{d}})\citenamefont {Wagner}, \citenamefont {Schwartzman-Nowik}, \citenamefont {Paiva}, \citenamefont {Te’eni}, \citenamefont {Ruiz-Molero}, \citenamefont {Barbosa}, \citenamefont {Cohen},\ and\ \citenamefont {Galvão}}]{wagner2024quantum}%
  \BibitemOpen
  \bibfield  {author} {\bibinfo {author} {\bibfnamefont {R.}~\bibnamefont {Wagner}}, \bibinfo {author} {\bibfnamefont {Z.}~\bibnamefont {Schwartzman-Nowik}}, \bibinfo {author} {\bibfnamefont {I.~L.}\ \bibnamefont {Paiva}}, \bibinfo {author} {\bibfnamefont {A.}~\bibnamefont {Te’eni}}, \bibinfo {author} {\bibfnamefont {A.}~\bibnamefont {Ruiz-Molero}}, \bibinfo {author} {\bibfnamefont {R.~S.}\ \bibnamefont {Barbosa}}, \bibinfo {author} {\bibfnamefont {E.}~\bibnamefont {Cohen}},\ and\ \bibinfo {author} {\bibfnamefont {E.~F.}\ \bibnamefont {Galvão}},\ }\bibfield  {title} {\bibinfo {title} {{Quantum circuits for measuring weak values, Kirkwood–Dirac quasiprobability distributions, and state spectra}},\ }\href {https://doi.org/10.1088/2058-9565/ad124c} {\bibfield  {journal} {\bibinfo  {journal} {Quantum Sci. Technol.}\ }\textbf {\bibinfo {volume} {9}},\ \bibinfo {pages} {015030} (\bibinfo {year} {2024}{\natexlab{d}})}\BibitemShut {NoStop}%
\bibitem [{\citenamefont {Gherardini}\ and\ \citenamefont {De~Chiara}(2024)}]{gherardini2024quasiprobabilities}%
  \BibitemOpen
  \bibfield  {author} {\bibinfo {author} {\bibfnamefont {S.}~\bibnamefont {Gherardini}}\ and\ \bibinfo {author} {\bibfnamefont {G.}~\bibnamefont {De~Chiara}},\ }\bibfield  {title} {\bibinfo {title} {{Quasiprobabilities in Quantum Thermodynamics and Many-Body Systems}},\ }\href {https://doi.org/10.1103/PRXQuantum.5.030201} {\bibfield  {journal} {\bibinfo  {journal} {PRX Quantum}\ }\textbf {\bibinfo {volume} {5}},\ \bibinfo {pages} {030201} (\bibinfo {year} {2024})}\BibitemShut {NoStop}%
\bibitem [{\citenamefont {Arvidsson-Shukur}\ \emph {et~al.}(2024)\citenamefont {Arvidsson-Shukur}, \citenamefont {Braasch~Jr}, \citenamefont {De~Bièvre}, \citenamefont {Dressel}, \citenamefont {Jordan}, \citenamefont {Langrenez}, \citenamefont {Lostaglio}, \citenamefont {Lundeen},\ and\ \citenamefont {Halpern}}]{arvidssonshukur2024properties}%
  \BibitemOpen
  \bibfield  {author} {\bibinfo {author} {\bibfnamefont {D.~R.~M.}\ \bibnamefont {Arvidsson-Shukur}}, \bibinfo {author} {\bibfnamefont {W.~F.}\ \bibnamefont {Braasch~Jr}}, \bibinfo {author} {\bibfnamefont {S.}~\bibnamefont {De~Bièvre}}, \bibinfo {author} {\bibfnamefont {J.}~\bibnamefont {Dressel}}, \bibinfo {author} {\bibfnamefont {A.~N.}\ \bibnamefont {Jordan}}, \bibinfo {author} {\bibfnamefont {C.}~\bibnamefont {Langrenez}}, \bibinfo {author} {\bibfnamefont {M.}~\bibnamefont {Lostaglio}}, \bibinfo {author} {\bibfnamefont {J.~S.}\ \bibnamefont {Lundeen}},\ and\ \bibinfo {author} {\bibfnamefont {N.~Y.}\ \bibnamefont {Halpern}},\ }\bibfield  {title} {\bibinfo {title} {Properties and applications of the {K}irkwood--{D}irac distribution},\ }\href {https://doi.org/10.1088/1367-2630/ada05d} {\bibfield  {journal} {\bibinfo  {journal} {New J. Phys.}\ }\textbf {\bibinfo {volume} {26}},\ \bibinfo {pages} {121201} (\bibinfo {year} {2024})}\BibitemShut {NoStop}%
\bibitem [{\citenamefont {Schmid}\ \emph {et~al.}(2024)\citenamefont {Schmid}, \citenamefont {Baldij\~ao}, \citenamefont {Y\ifmmode~\bar{\imath}\else \={\i}\fi{}ng}, \citenamefont {Wagner},\ and\ \citenamefont {Selby}}]{schmid2024kirkwooddirac}%
  \BibitemOpen
  \bibfield  {author} {\bibinfo {author} {\bibfnamefont {D.}~\bibnamefont {Schmid}}, \bibinfo {author} {\bibfnamefont {R.~D.}\ \bibnamefont {Baldij\~ao}}, \bibinfo {author} {\bibfnamefont {Y.}~\bibnamefont {Y\ifmmode~\bar{\imath}\else \={\i}\fi{}ng}}, \bibinfo {author} {\bibfnamefont {R.}~\bibnamefont {Wagner}},\ and\ \bibinfo {author} {\bibfnamefont {J.~H.}\ \bibnamefont {Selby}},\ }\bibfield  {title} {\bibinfo {title} {Kirkwood--{D}irac representations beyond quantum states and their relation to noncontextuality},\ }\href {https://doi.org/10.1103/PhysRevA.110.052206} {\bibfield  {journal} {\bibinfo  {journal} {Phys. Rev. A}\ }\textbf {\bibinfo {volume} {110}},\ \bibinfo {pages} {052206} (\bibinfo {year} {2024})}\BibitemShut {NoStop}%
\bibitem [{\citenamefont {Liu}\ and\ \citenamefont {Cheng}(2025)}]{liu2025boundarykirkwooddiracquasiprobability}%
  \BibitemOpen
  \bibfield  {author} {\bibinfo {author} {\bibfnamefont {L.}~\bibnamefont {Liu}}\ and\ \bibinfo {author} {\bibfnamefont {S.}~\bibnamefont {Cheng}},\ }\href {https://arxiv.org/abs/2504.09238} {\bibinfo {title} {{The boundary of Kirkwood-Dirac quasiprobability}}} (\bibinfo {year} {2025}),\ \Eprint {https://arxiv.org/abs/2504.09238} {arXiv:2504.09238 [quant-ph]} \BibitemShut {NoStop}%
\bibitem [{\citenamefont {Alves}\ \emph {et~al.}(2003)\citenamefont {Alves}, \citenamefont {Horodecki}, \citenamefont {Oi}, \citenamefont {Kwek},\ and\ \citenamefont {Ekert}}]{alves2003direct}%
  \BibitemOpen
  \bibfield  {author} {\bibinfo {author} {\bibfnamefont {C.~M.}\ \bibnamefont {Alves}}, \bibinfo {author} {\bibfnamefont {P.}~\bibnamefont {Horodecki}}, \bibinfo {author} {\bibfnamefont {D.~K.~L.}\ \bibnamefont {Oi}}, \bibinfo {author} {\bibfnamefont {L.~C.}\ \bibnamefont {Kwek}},\ and\ \bibinfo {author} {\bibfnamefont {A.~K.}\ \bibnamefont {Ekert}},\ }\bibfield  {title} {\bibinfo {title} {Direct estimation of functionals of density operators by local operations and classical communication},\ }\href {https://doi.org/10.1103/PhysRevA.68.032306} {\bibfield  {journal} {\bibinfo  {journal} {Phys. Rev. A}\ }\textbf {\bibinfo {volume} {68}},\ \bibinfo {pages} {032306} (\bibinfo {year} {2003})}\BibitemShut {NoStop}%
\bibitem [{\citenamefont {Brun}(2004)}]{brun2004measuring}%
  \BibitemOpen
  \bibfield  {author} {\bibinfo {author} {\bibfnamefont {T.}~\bibnamefont {Brun}},\ }\bibfield  {title} {\bibinfo {title} {{Measuring polynomial functions of states}},\ }\href {https://doi.org/10.26421/qic4.5-6} {\bibfield  {journal} {\bibinfo  {journal} {Quantum Inf. Comput.}\ }\textbf {\bibinfo {volume} {4}},\ \bibinfo {pages} {401} (\bibinfo {year} {2004})}\BibitemShut {NoStop}%
\bibitem [{\citenamefont {Leifer}\ \emph {et~al.}(2004)\citenamefont {Leifer}, \citenamefont {Linden},\ and\ \citenamefont {Winter}}]{leifer2004measuring}%
  \BibitemOpen
  \bibfield  {author} {\bibinfo {author} {\bibfnamefont {M.~S.}\ \bibnamefont {Leifer}}, \bibinfo {author} {\bibfnamefont {N.}~\bibnamefont {Linden}},\ and\ \bibinfo {author} {\bibfnamefont {A.}~\bibnamefont {Winter}},\ }\bibfield  {title} {\bibinfo {title} {Measuring polynomial invariants of multiparty quantum states},\ }\href {https://doi.org/10.1103/PhysRevA.69.052304} {\bibfield  {journal} {\bibinfo  {journal} {Phys. Rev. A}\ }\textbf {\bibinfo {volume} {69}},\ \bibinfo {pages} {052304} (\bibinfo {year} {2004})}\BibitemShut {NoStop}%
\bibitem [{\citenamefont {van Enk}\ and\ \citenamefont {Beenakker}(2012)}]{vanenk2012measuring}%
  \BibitemOpen
  \bibfield  {author} {\bibinfo {author} {\bibfnamefont {S.~J.}\ \bibnamefont {van Enk}}\ and\ \bibinfo {author} {\bibfnamefont {C.~W.~J.}\ \bibnamefont {Beenakker}},\ }\bibfield  {title} {\bibinfo {title} {{Measuring $\mathrm{Tr}{\ensuremath{\rho}}^{n}$ on Single Copies of $\ensuremath{\rho}$ Using Random Measurements}},\ }\href {https://doi.org/10.1103/PhysRevLett.108.110503} {\bibfield  {journal} {\bibinfo  {journal} {Phys. Rev. Lett.}\ }\textbf {\bibinfo {volume} {108}},\ \bibinfo {pages} {110503} (\bibinfo {year} {2012})}\BibitemShut {NoStop}%
\bibitem [{\citenamefont {Tanaka}\ \emph {et~al.}(2014)\citenamefont {Tanaka}, \citenamefont {Ota}, \citenamefont {Kanazawa}, \citenamefont {Kimura}, \citenamefont {Nakazato},\ and\ \citenamefont {Nori}}]{tanaka2014determining}%
  \BibitemOpen
  \bibfield  {author} {\bibinfo {author} {\bibfnamefont {T.}~\bibnamefont {Tanaka}}, \bibinfo {author} {\bibfnamefont {Y.}~\bibnamefont {Ota}}, \bibinfo {author} {\bibfnamefont {M.}~\bibnamefont {Kanazawa}}, \bibinfo {author} {\bibfnamefont {G.}~\bibnamefont {Kimura}}, \bibinfo {author} {\bibfnamefont {H.}~\bibnamefont {Nakazato}},\ and\ \bibinfo {author} {\bibfnamefont {F.}~\bibnamefont {Nori}},\ }\bibfield  {title} {\bibinfo {title} {Determining eigenvalues of a density matrix with minimal information in a single experimental setting},\ }\href {https://doi.org/10.1103/PhysRevA.89.012117} {\bibfield  {journal} {\bibinfo  {journal} {Phys. Rev. A}\ }\textbf {\bibinfo {volume} {89}},\ \bibinfo {pages} {012117} (\bibinfo {year} {2014})}\BibitemShut {NoStop}%
\bibitem [{\citenamefont {Shin}\ \emph {et~al.}(2024)\citenamefont {Shin}, \citenamefont {Lee}, \citenamefont {Lee},\ and\ \citenamefont {Jeong}}]{shin2024rankneedestimatingtrace}%
  \BibitemOpen
  \bibfield  {author} {\bibinfo {author} {\bibfnamefont {M.}~\bibnamefont {Shin}}, \bibinfo {author} {\bibfnamefont {J.}~\bibnamefont {Lee}}, \bibinfo {author} {\bibfnamefont {S.}~\bibnamefont {Lee}},\ and\ \bibinfo {author} {\bibfnamefont {K.}~\bibnamefont {Jeong}},\ }\href {https://arxiv.org/abs/2408.00314} {\bibinfo {title} {Rank is all you need: Estimating the trace of powers of density matrices}} (\bibinfo {year} {2024}),\ \Eprint {https://arxiv.org/abs/2408.00314} {arXiv:2408.00314 [quant-ph]} \BibitemShut {NoStop}%
\bibitem [{\citenamefont {Johri}\ \emph {et~al.}(2017)\citenamefont {Johri}, \citenamefont {Steiger},\ and\ \citenamefont {Troyer}}]{johri2017entanglement}%
  \BibitemOpen
  \bibfield  {author} {\bibinfo {author} {\bibfnamefont {S.}~\bibnamefont {Johri}}, \bibinfo {author} {\bibfnamefont {D.~S.}\ \bibnamefont {Steiger}},\ and\ \bibinfo {author} {\bibfnamefont {M.}~\bibnamefont {Troyer}},\ }\bibfield  {title} {\bibinfo {title} {Entanglement spectroscopy on a quantum computer},\ }\href {https://doi.org/10.1103/PhysRevB.96.195136} {\bibfield  {journal} {\bibinfo  {journal} {Phys. Rev. B}\ }\textbf {\bibinfo {volume} {96}},\ \bibinfo {pages} {195136} (\bibinfo {year} {2017})}\BibitemShut {NoStop}%
\bibitem [{\citenamefont {Turkeshi}\ \emph {et~al.}(2023)\citenamefont {Turkeshi}, \citenamefont {Schir\`o},\ and\ \citenamefont {Sierant}}]{turkeshi2023measuring}%
  \BibitemOpen
  \bibfield  {author} {\bibinfo {author} {\bibfnamefont {X.}~\bibnamefont {Turkeshi}}, \bibinfo {author} {\bibfnamefont {M.}~\bibnamefont {Schir\`o}},\ and\ \bibinfo {author} {\bibfnamefont {P.}~\bibnamefont {Sierant}},\ }\bibfield  {title} {\bibinfo {title} {Measuring nonstabilizerness via multifractal flatness},\ }\href {https://doi.org/10.1103/PhysRevA.108.042408} {\bibfield  {journal} {\bibinfo  {journal} {Phys. Rev. A}\ }\textbf {\bibinfo {volume} {108}},\ \bibinfo {pages} {042408} (\bibinfo {year} {2023})}\BibitemShut {NoStop}%
\bibitem [{\citenamefont {Tirrito}\ \emph {et~al.}(2024)\citenamefont {Tirrito}, \citenamefont {Tarabunga}, \citenamefont {Lami}, \citenamefont {Chanda}, \citenamefont {Leone}, \citenamefont {Oliviero}, \citenamefont {Dalmonte}, \citenamefont {Collura},\ and\ \citenamefont {Hamma}}]{tirrito2024quantifying}%
  \BibitemOpen
  \bibfield  {author} {\bibinfo {author} {\bibfnamefont {E.}~\bibnamefont {Tirrito}}, \bibinfo {author} {\bibfnamefont {P.~S.}\ \bibnamefont {Tarabunga}}, \bibinfo {author} {\bibfnamefont {G.}~\bibnamefont {Lami}}, \bibinfo {author} {\bibfnamefont {T.}~\bibnamefont {Chanda}}, \bibinfo {author} {\bibfnamefont {L.}~\bibnamefont {Leone}}, \bibinfo {author} {\bibfnamefont {S.~F.~E.}\ \bibnamefont {Oliviero}}, \bibinfo {author} {\bibfnamefont {M.}~\bibnamefont {Dalmonte}}, \bibinfo {author} {\bibfnamefont {M.}~\bibnamefont {Collura}},\ and\ \bibinfo {author} {\bibfnamefont {A.}~\bibnamefont {Hamma}},\ }\bibfield  {title} {\bibinfo {title} {Quantifying nonstabilizerness through entanglement spectrum flatness},\ }\href {https://doi.org/10.1103/PhysRevA.109.L040401} {\bibfield  {journal} {\bibinfo  {journal} {Phys. Rev. A}\ }\textbf {\bibinfo {volume} {109}},\ \bibinfo {pages} {L040401} (\bibinfo {year} {2024})}\BibitemShut {NoStop}%
\bibitem [{\citenamefont {De~Chiara}\ \emph {et~al.}(2012)\citenamefont {De~Chiara}, \citenamefont {Lepori}, \citenamefont {Lewenstein},\ and\ \citenamefont {Sanpera}}]{dechiara2012entanglement}%
  \BibitemOpen
  \bibfield  {author} {\bibinfo {author} {\bibfnamefont {G.}~\bibnamefont {De~Chiara}}, \bibinfo {author} {\bibfnamefont {L.}~\bibnamefont {Lepori}}, \bibinfo {author} {\bibfnamefont {M.}~\bibnamefont {Lewenstein}},\ and\ \bibinfo {author} {\bibfnamefont {A.}~\bibnamefont {Sanpera}},\ }\bibfield  {title} {\bibinfo {title} {{Entanglement Spectrum, Critical Exponents, and Order Parameters in Quantum Spin Chains}},\ }\href {https://doi.org/10.1103/PhysRevLett.109.237208} {\bibfield  {journal} {\bibinfo  {journal} {Phys. Rev. Lett.}\ }\textbf {\bibinfo {volume} {109}},\ \bibinfo {pages} {237208} (\bibinfo {year} {2012})}\BibitemShut {NoStop}%
\bibitem [{\citenamefont {Chamon}\ \emph {et~al.}(2014)\citenamefont {Chamon}, \citenamefont {Hamma},\ and\ \citenamefont {Mucciolo}}]{chamon2014emergent}%
  \BibitemOpen
  \bibfield  {author} {\bibinfo {author} {\bibfnamefont {C.}~\bibnamefont {Chamon}}, \bibinfo {author} {\bibfnamefont {A.}~\bibnamefont {Hamma}},\ and\ \bibinfo {author} {\bibfnamefont {E.~R.}\ \bibnamefont {Mucciolo}},\ }\bibfield  {title} {\bibinfo {title} {{Emergent Irreversibility and Entanglement Spectrum Statistics}},\ }\href {https://doi.org/10.1103/PhysRevLett.112.240501} {\bibfield  {journal} {\bibinfo  {journal} {Phys. Rev. Lett.}\ }\textbf {\bibinfo {volume} {112}},\ \bibinfo {pages} {240501} (\bibinfo {year} {2014})}\BibitemShut {NoStop}%
\bibitem [{\citenamefont {Koczor}(2021)}]{koczor2021exponential}%
  \BibitemOpen
  \bibfield  {author} {\bibinfo {author} {\bibfnamefont {B.}~\bibnamefont {Koczor}},\ }\bibfield  {title} {\bibinfo {title} {{Exponential Error Suppression for Near-Term Quantum Devices}},\ }\href {https://doi.org/10.1103/PhysRevX.11.031057} {\bibfield  {journal} {\bibinfo  {journal} {Phys. Rev. X}\ }\textbf {\bibinfo {volume} {11}},\ \bibinfo {pages} {031057} (\bibinfo {year} {2021})}\BibitemShut {NoStop}%
\bibitem [{\citenamefont {Huggins}\ \emph {et~al.}(2021)\citenamefont {Huggins}, \citenamefont {McArdle}, \citenamefont {O'Brien}, \citenamefont {Lee}, \citenamefont {Rubin}, \citenamefont {Boixo}, \citenamefont {Whaley}, \citenamefont {Babbush},\ and\ \citenamefont {McClean}}]{huggins2021virtual}%
  \BibitemOpen
  \bibfield  {author} {\bibinfo {author} {\bibfnamefont {W.~J.}\ \bibnamefont {Huggins}}, \bibinfo {author} {\bibfnamefont {S.}~\bibnamefont {McArdle}}, \bibinfo {author} {\bibfnamefont {T.~E.}\ \bibnamefont {O'Brien}}, \bibinfo {author} {\bibfnamefont {J.}~\bibnamefont {Lee}}, \bibinfo {author} {\bibfnamefont {N.~C.}\ \bibnamefont {Rubin}}, \bibinfo {author} {\bibfnamefont {S.}~\bibnamefont {Boixo}}, \bibinfo {author} {\bibfnamefont {K.~B.}\ \bibnamefont {Whaley}}, \bibinfo {author} {\bibfnamefont {R.}~\bibnamefont {Babbush}},\ and\ \bibinfo {author} {\bibfnamefont {J.~R.}\ \bibnamefont {McClean}},\ }\bibfield  {title} {\bibinfo {title} {Virtual distillation for quantum error mitigation},\ }\href {https://doi.org/10.1103/PhysRevX.11.041036} {\bibfield  {journal} {\bibinfo  {journal} {Phys. Rev. X}\ }\textbf {\bibinfo {volume} {11}},\ \bibinfo {pages} {041036} (\bibinfo {year} {2021})}\BibitemShut {NoStop}%
\bibitem [{\citenamefont {Wang}\ \emph {et~al.}(2021)\citenamefont {Wang}, \citenamefont {Li},\ and\ \citenamefont {Wang}}]{wang2021variational}%
  \BibitemOpen
  \bibfield  {author} {\bibinfo {author} {\bibfnamefont {Y.}~\bibnamefont {Wang}}, \bibinfo {author} {\bibfnamefont {G.}~\bibnamefont {Li}},\ and\ \bibinfo {author} {\bibfnamefont {X.}~\bibnamefont {Wang}},\ }\bibfield  {title} {\bibinfo {title} {{Variational Quantum Gibbs State Preparation with a Truncated Taylor Series}},\ }\href {https://doi.org/10.1103/PhysRevApplied.16.054035} {\bibfield  {journal} {\bibinfo  {journal} {Phys. Rev. Appl.}\ }\textbf {\bibinfo {volume} {16}},\ \bibinfo {pages} {054035} (\bibinfo {year} {2021})}\BibitemShut {NoStop}%
\bibitem [{\citenamefont {Cotler}\ \emph {et~al.}(2019)\citenamefont {Cotler}, \citenamefont {Choi}, \citenamefont {Lukin}, \citenamefont {Gharibyan}, \citenamefont {Grover}, \citenamefont {Tai}, \citenamefont {Rispoli}, \citenamefont {Schittko}, \citenamefont {Preiss}, \citenamefont {Kaufman}, \citenamefont {Greiner}, \citenamefont {Pichler},\ and\ \citenamefont {Hayden}}]{cotler2019cooling}%
  \BibitemOpen
  \bibfield  {author} {\bibinfo {author} {\bibfnamefont {J.}~\bibnamefont {Cotler}}, \bibinfo {author} {\bibfnamefont {S.}~\bibnamefont {Choi}}, \bibinfo {author} {\bibfnamefont {A.}~\bibnamefont {Lukin}}, \bibinfo {author} {\bibfnamefont {H.}~\bibnamefont {Gharibyan}}, \bibinfo {author} {\bibfnamefont {T.}~\bibnamefont {Grover}}, \bibinfo {author} {\bibfnamefont {M.~E.}\ \bibnamefont {Tai}}, \bibinfo {author} {\bibfnamefont {M.}~\bibnamefont {Rispoli}}, \bibinfo {author} {\bibfnamefont {R.}~\bibnamefont {Schittko}}, \bibinfo {author} {\bibfnamefont {P.~M.}\ \bibnamefont {Preiss}}, \bibinfo {author} {\bibfnamefont {A.~M.}\ \bibnamefont {Kaufman}}, \bibinfo {author} {\bibfnamefont {M.}~\bibnamefont {Greiner}}, \bibinfo {author} {\bibfnamefont {H.}~\bibnamefont {Pichler}},\ and\ \bibinfo {author} {\bibfnamefont {P.}~\bibnamefont {Hayden}},\ }\bibfield  {title} {\bibinfo {title} {Quantum virtual cooling},\ }\href {https://doi.org/10.1103/PhysRevX.9.031013} {\bibfield  {journal} {\bibinfo  {journal} {Phys. Rev.
  X}\ }\textbf {\bibinfo {volume} {9}},\ \bibinfo {pages} {031013} (\bibinfo {year} {2019})}\BibitemShut {NoStop}%
\bibitem [{\citenamefont {Gao}\ \emph {et~al.}(2023)\citenamefont {Gao}, \citenamefont {Li}, \citenamefont {Mishra}, \citenamefont {Yan}, \citenamefont {Simonov},\ and\ \citenamefont {Chiribella}}]{gao2023measuring}%
  \BibitemOpen
  \bibfield  {author} {\bibinfo {author} {\bibfnamefont {N.}~\bibnamefont {Gao}}, \bibinfo {author} {\bibfnamefont {D.}~\bibnamefont {Li}}, \bibinfo {author} {\bibfnamefont {A.}~\bibnamefont {Mishra}}, \bibinfo {author} {\bibfnamefont {J.}~\bibnamefont {Yan}}, \bibinfo {author} {\bibfnamefont {K.}~\bibnamefont {Simonov}},\ and\ \bibinfo {author} {\bibfnamefont {G.}~\bibnamefont {Chiribella}},\ }\bibfield  {title} {\bibinfo {title} {{Measuring Incompatibility and Clustering Quantum Observables with a Quantum Switch}},\ }\href {https://doi.org/10.1103/PhysRevLett.130.170201} {\bibfield  {journal} {\bibinfo  {journal} {Phys. Rev. Lett.}\ }\textbf {\bibinfo {volume} {130}},\ \bibinfo {pages} {170201} (\bibinfo {year} {2023})}\BibitemShut {NoStop}%
\bibitem [{\citenamefont {Yunger~Halpern}\ \emph {et~al.}(2018)\citenamefont {Yunger~Halpern}, \citenamefont {Swingle},\ and\ \citenamefont {Dressel}}]{halpern2018quasiprobability}%
  \BibitemOpen
  \bibfield  {author} {\bibinfo {author} {\bibfnamefont {N.}~\bibnamefont {Yunger~Halpern}}, \bibinfo {author} {\bibfnamefont {B.}~\bibnamefont {Swingle}},\ and\ \bibinfo {author} {\bibfnamefont {J.}~\bibnamefont {Dressel}},\ }\bibfield  {title} {\bibinfo {title} {Quasiprobability behind the out-of-time-ordered correlator},\ }\href {https://doi.org/10.1103/PhysRevA.97.042105} {\bibfield  {journal} {\bibinfo  {journal} {Phys. Rev. A}\ }\textbf {\bibinfo {volume} {97}},\ \bibinfo {pages} {042105} (\bibinfo {year} {2018})}\BibitemShut {NoStop}%
\bibitem [{\citenamefont {Gonz\'alez~Alonso}\ \emph {et~al.}(2019)\citenamefont {Gonz\'alez~Alonso}, \citenamefont {Yunger~Halpern},\ and\ \citenamefont {Dressel}}]{gonzales2019otoc}%
  \BibitemOpen
  \bibfield  {author} {\bibinfo {author} {\bibfnamefont {J.~R.}\ \bibnamefont {Gonz\'alez~Alonso}}, \bibinfo {author} {\bibfnamefont {N.}~\bibnamefont {Yunger~Halpern}},\ and\ \bibinfo {author} {\bibfnamefont {J.}~\bibnamefont {Dressel}},\ }\bibfield  {title} {\bibinfo {title} {{Out-of-Time-Ordered-Correlator Quasiprobabilities Robustly Witness Scrambling}},\ }\href {https://doi.org/10.1103/PhysRevLett.122.040404} {\bibfield  {journal} {\bibinfo  {journal} {Phys. Rev. Lett.}\ }\textbf {\bibinfo {volume} {122}},\ \bibinfo {pages} {040404} (\bibinfo {year} {2019})}\BibitemShut {NoStop}%
\bibitem [{\citenamefont {Alonso}\ \emph {et~al.}(2022)\citenamefont {Alonso}, \citenamefont {Shammah}, \citenamefont {Ahmed}, \citenamefont {Nori},\ and\ \citenamefont {Dressel}}]{gonzales2022diagnosing}%
  \BibitemOpen
  \bibfield  {author} {\bibinfo {author} {\bibfnamefont {J.~R.~G.}\ \bibnamefont {Alonso}}, \bibinfo {author} {\bibfnamefont {N.}~\bibnamefont {Shammah}}, \bibinfo {author} {\bibfnamefont {S.}~\bibnamefont {Ahmed}}, \bibinfo {author} {\bibfnamefont {F.}~\bibnamefont {Nori}},\ and\ \bibinfo {author} {\bibfnamefont {J.}~\bibnamefont {Dressel}},\ }\href {https://doi.org/https://doi.org/10.48550/arXiv.2201.08175} {\bibinfo {title} {Diagnosing quantum chaos with out-of-time-ordered-correlator quasiprobability in the kicked-top model}} (\bibinfo {year} {2022}),\ \Eprint {https://arxiv.org/abs/2201.08175} {arXiv:2201.08175 [quant-ph]} \BibitemShut {NoStop}%
\bibitem [{\citenamefont {Mitchison}\ \emph {et~al.}(2007)\citenamefont {Mitchison}, \citenamefont {Jozsa},\ and\ \citenamefont {Popescu}}]{mitchison2007sequential}%
  \BibitemOpen
  \bibfield  {author} {\bibinfo {author} {\bibfnamefont {G.}~\bibnamefont {Mitchison}}, \bibinfo {author} {\bibfnamefont {R.}~\bibnamefont {Jozsa}},\ and\ \bibinfo {author} {\bibfnamefont {S.}~\bibnamefont {Popescu}},\ }\bibfield  {title} {\bibinfo {title} {Sequential weak measurement},\ }\href {https://doi.org/10.1103/PhysRevA.76.062105} {\bibfield  {journal} {\bibinfo  {journal} {Phys. Rev. A}\ }\textbf {\bibinfo {volume} {76}},\ \bibinfo {pages} {062105} (\bibinfo {year} {2007})}\BibitemShut {NoStop}%
\bibitem [{\citenamefont {Pancharatnam}(1956)}]{pancharatnam1956generalized}%
  \BibitemOpen
  \bibfield  {author} {\bibinfo {author} {\bibfnamefont {S.}~\bibnamefont {Pancharatnam}},\ }\bibfield  {title} {\bibinfo {title} {Generalized theory of interference, and its applications: Part {I}. {C}oherent pencils},\ }\href {https://doi.org/10.1007/bf03046050} {\bibfield  {journal} {\bibinfo  {journal} {Proc. Indian Acad. Sci.}\ }\textbf {\bibinfo {volume} {44}},\ \bibinfo {pages} {247} (\bibinfo {year} {1956})}\BibitemShut {NoStop}%
\bibitem [{\citenamefont {Arvind}\ \emph {et~al.}(1997)\citenamefont {Arvind}, \citenamefont {Mallesh},\ and\ \citenamefont {Mukunda}}]{arvind1997generalized}%
  \BibitemOpen
  \bibfield  {author} {\bibinfo {author} {\bibnamefont {Arvind}}, \bibinfo {author} {\bibfnamefont {K.~S.}\ \bibnamefont {Mallesh}},\ and\ \bibinfo {author} {\bibfnamefont {N.}~\bibnamefont {Mukunda}},\ }\bibfield  {title} {\bibinfo {title} {A generalized {P}ancharatnam geometric phase formula for three-level quantum systems},\ }\href {https://doi.org/10.1088/0305-4470/30/7/021} {\bibfield  {journal} {\bibinfo  {journal} {J. Phys. A: Math. Gen.}\ }\textbf {\bibinfo {volume} {30}},\ \bibinfo {pages} {2417} (\bibinfo {year} {1997})}\BibitemShut {NoStop}%
\bibitem [{\citenamefont {Wilczek}\ and\ \citenamefont {Shapere}(1989)}]{wilczek1989geometric}%
  \BibitemOpen
  \bibfield  {author} {\bibinfo {author} {\bibfnamefont {F.}~\bibnamefont {Wilczek}}\ and\ \bibinfo {author} {\bibfnamefont {A.}~\bibnamefont {Shapere}},\ }\href {https://doi.org/10.1142/0613} {\emph {\bibinfo {title} {{Geometric Phases in Physics}}}},\ Advanced Series in Mathematical Physics\ (\bibinfo  {publisher} {WORLD SCIENTIFIC},\ \bibinfo {year} {1989})\BibitemShut {NoStop}%
\bibitem [{\citenamefont {Chruściński}\ and\ \citenamefont {Jamiołkowski}(2004)}]{chruciski2004geometric}%
  \BibitemOpen
  \bibfield  {author} {\bibinfo {author} {\bibfnamefont {D.}~\bibnamefont {Chruściński}}\ and\ \bibinfo {author} {\bibfnamefont {A.}~\bibnamefont {Jamiołkowski}},\ }\href {https://doi.org/10.1007/978-0-8176-8176-0} {\emph {\bibinfo {title} {{Geometric Phases in Classical and Quantum Mechanics}}}}\ (\bibinfo  {publisher} {Birkhäuser Boston},\ \bibinfo {year} {2004})\BibitemShut {NoStop}%
\bibitem [{\citenamefont {Fernandes}\ \emph {et~al.}(2024)\citenamefont {Fernandes}, \citenamefont {Wagner}, \citenamefont {Novo},\ and\ \citenamefont {Galv\~ao}}]{fernandes2024unitaryinvariant}%
  \BibitemOpen
  \bibfield  {author} {\bibinfo {author} {\bibfnamefont {C.}~\bibnamefont {Fernandes}}, \bibinfo {author} {\bibfnamefont {R.}~\bibnamefont {Wagner}}, \bibinfo {author} {\bibfnamefont {L.}~\bibnamefont {Novo}},\ and\ \bibinfo {author} {\bibfnamefont {E.~F.}\ \bibnamefont {Galv\~ao}},\ }\bibfield  {title} {\bibinfo {title} {{Unitary-Invariant Witnesses of Quantum Imaginarity}},\ }\href {https://doi.org/10.1103/PhysRevLett.133.190201} {\bibfield  {journal} {\bibinfo  {journal} {Phys. Rev. Lett.}\ }\textbf {\bibinfo {volume} {133}},\ \bibinfo {pages} {190201} (\bibinfo {year} {2024})}\BibitemShut {NoStop}%
\bibitem [{\citenamefont {Berry}(1984)}]{berry1984quantal}%
  \BibitemOpen
  \bibfield  {author} {\bibinfo {author} {\bibfnamefont {M.~V.}\ \bibnamefont {Berry}},\ }\bibfield  {title} {\bibinfo {title} {Quantal phase factors accompanying adiabatic changes},\ }\href {https://doi.org/https://doi.org/10.1098/rspa.1984.0023} {\bibfield  {journal} {\bibinfo  {journal} {Proc. R. Soc. Lond. A}\ }\textbf {\bibinfo {volume} {392}},\ \bibinfo {pages} {45} (\bibinfo {year} {1984})}\BibitemShut {NoStop}%
\bibitem [{\citenamefont {Avdoshkin}\ and\ \citenamefont {Popov}(2023)}]{avdoshkin2023extrinsic}%
  \BibitemOpen
  \bibfield  {author} {\bibinfo {author} {\bibfnamefont {A.}~\bibnamefont {Avdoshkin}}\ and\ \bibinfo {author} {\bibfnamefont {F.~K.}\ \bibnamefont {Popov}},\ }\bibfield  {title} {\bibinfo {title} {Extrinsic geometry of quantum states},\ }\href {https://doi.org/10.1103/PhysRevB.107.245136} {\bibfield  {journal} {\bibinfo  {journal} {Phys. Rev. B}\ }\textbf {\bibinfo {volume} {107}},\ \bibinfo {pages} {245136} (\bibinfo {year} {2023})}\BibitemShut {NoStop}%
\bibitem [{\citenamefont {Rabei}\ \emph {et~al.}(1999)\citenamefont {Rabei}, \citenamefont {Arvind}, \citenamefont {Mukunda},\ and\ \citenamefont {Simon}}]{rabei1999Bargmann}%
  \BibitemOpen
  \bibfield  {author} {\bibinfo {author} {\bibfnamefont {E.~M.}\ \bibnamefont {Rabei}}, \bibinfo {author} {\bibnamefont {Arvind}}, \bibinfo {author} {\bibfnamefont {N.}~\bibnamefont {Mukunda}},\ and\ \bibinfo {author} {\bibfnamefont {R.}~\bibnamefont {Simon}},\ }\bibfield  {title} {\bibinfo {title} {Bargmann invariants and geometric phases: A generalized connection},\ }\href {https://doi.org/10.1103/PhysRevA.60.3397} {\bibfield  {journal} {\bibinfo  {journal} {Phys. Rev. A}\ }\textbf {\bibinfo {volume} {60}},\ \bibinfo {pages} {3397} (\bibinfo {year} {1999})}\BibitemShut {NoStop}%
\bibitem [{\citenamefont {Quek}\ \emph {et~al.}(2024)\citenamefont {Quek}, \citenamefont {Kaur},\ and\ \citenamefont {Wilde}}]{quek2024multivariatetrace}%
  \BibitemOpen
  \bibfield  {author} {\bibinfo {author} {\bibfnamefont {Y.}~\bibnamefont {Quek}}, \bibinfo {author} {\bibfnamefont {E.}~\bibnamefont {Kaur}},\ and\ \bibinfo {author} {\bibfnamefont {M.~M.}\ \bibnamefont {Wilde}},\ }\bibfield  {title} {\bibinfo {title} {Multivariate trace estimation in constant quantum depth},\ }\href {https://doi.org/10.22331/q-2024-01-10-1220} {\bibfield  {journal} {\bibinfo  {journal} {{Quantum}}\ }\textbf {\bibinfo {volume} {8}},\ \bibinfo {pages} {1220} (\bibinfo {year} {2024})}\BibitemShut {NoStop}%
\bibitem [{\citenamefont {Chiribella}\ \emph {et~al.}(2024)\citenamefont {Chiribella}, \citenamefont {Simonov},\ and\ \citenamefont {Zhao}}]{chiribella2024dimensionindependentweakvalueestimation}%
  \BibitemOpen
  \bibfield  {author} {\bibinfo {author} {\bibfnamefont {G.}~\bibnamefont {Chiribella}}, \bibinfo {author} {\bibfnamefont {K.}~\bibnamefont {Simonov}},\ and\ \bibinfo {author} {\bibfnamefont {X.}~\bibnamefont {Zhao}},\ }\bibfield  {title} {\bibinfo {title} {Dimension-independent weak value estimation via controlled {SWAP} operations},\ }\href {https://doi.org/10.1103/PhysRevResearch.6.043043} {\bibfield  {journal} {\bibinfo  {journal} {Phys. Rev. Research}\ }\textbf {\bibinfo {volume} {6}},\ \bibinfo {pages} {043043} (\bibinfo {year} {2024})}\BibitemShut {NoStop}%
\bibitem [{\citenamefont {Garcia-Escartin}\ and\ \citenamefont {Chamorro-Posada}(2013)}]{garcia2013swap}%
  \BibitemOpen
  \bibfield  {author} {\bibinfo {author} {\bibfnamefont {J.~C.}\ \bibnamefont {Garcia-Escartin}}\ and\ \bibinfo {author} {\bibfnamefont {P.}~\bibnamefont {Chamorro-Posada}},\ }\bibfield  {title} {\bibinfo {title} {swap test and {H}ong-{O}u-{M}andel effect are equivalent},\ }\href {https://doi.org/10.1103/PhysRevA.87.052330} {\bibfield  {journal} {\bibinfo  {journal} {Phys. Rev. A}\ }\textbf {\bibinfo {volume} {87}},\ \bibinfo {pages} {052330} (\bibinfo {year} {2013})}\BibitemShut {NoStop}%
\bibitem [{\citenamefont {Barenco}\ \emph {et~al.}(1997)\citenamefont {Barenco}, \citenamefont {Berthiaume}, \citenamefont {Deutsch}, \citenamefont {Ekert}, \citenamefont {Jozsa},\ and\ \citenamefont {Macchiavello}}]{barenco1997stabilization}%
  \BibitemOpen
  \bibfield  {author} {\bibinfo {author} {\bibfnamefont {A.}~\bibnamefont {Barenco}}, \bibinfo {author} {\bibfnamefont {A.}~\bibnamefont {Berthiaume}}, \bibinfo {author} {\bibfnamefont {D.}~\bibnamefont {Deutsch}}, \bibinfo {author} {\bibfnamefont {A.}~\bibnamefont {Ekert}}, \bibinfo {author} {\bibfnamefont {R.}~\bibnamefont {Jozsa}},\ and\ \bibinfo {author} {\bibfnamefont {C.}~\bibnamefont {Macchiavello}},\ }\bibfield  {title} {\bibinfo {title} {{Stabilization of Quantum Computations by Symmetrization}},\ }\href {https://doi.org/10.1137/s0097539796302452} {\bibfield  {journal} {\bibinfo  {journal} {SIAM J. Comput.}\ }\textbf {\bibinfo {volume} {26}},\ \bibinfo {pages} {1541–1557} (\bibinfo {year} {1997})}\BibitemShut {NoStop}%
\bibitem [{\citenamefont {Buhrman}\ \emph {et~al.}(2001)\citenamefont {Buhrman}, \citenamefont {Cleve}, \citenamefont {Watrous},\ and\ \citenamefont {de~Wolf}}]{buhrman2001quantumfingerprinting}%
  \BibitemOpen
  \bibfield  {author} {\bibinfo {author} {\bibfnamefont {H.}~\bibnamefont {Buhrman}}, \bibinfo {author} {\bibfnamefont {R.}~\bibnamefont {Cleve}}, \bibinfo {author} {\bibfnamefont {J.}~\bibnamefont {Watrous}},\ and\ \bibinfo {author} {\bibfnamefont {R.}~\bibnamefont {de~Wolf}},\ }\bibfield  {title} {\bibinfo {title} {{Quantum Fingerprinting}},\ }\href {https://doi.org/10.1103/PhysRevLett.87.167902} {\bibfield  {journal} {\bibinfo  {journal} {Phys. Rev. Lett.}\ }\textbf {\bibinfo {volume} {87}},\ \bibinfo {pages} {167902} (\bibinfo {year} {2001})}\BibitemShut {NoStop}%
\bibitem [{\citenamefont {Aharonov}\ \emph {et~al.}(2008)\citenamefont {Aharonov}, \citenamefont {Jones},\ and\ \citenamefont {Landau}}]{aaronov2008polynomial}%
  \BibitemOpen
  \bibfield  {author} {\bibinfo {author} {\bibfnamefont {D.}~\bibnamefont {Aharonov}}, \bibinfo {author} {\bibfnamefont {V.}~\bibnamefont {Jones}},\ and\ \bibinfo {author} {\bibfnamefont {Z.}~\bibnamefont {Landau}},\ }\bibfield  {title} {\bibinfo {title} {A {P}olynomial {Q}uantum {A}lgorithm for {A}pproximating the {J}ones {P}olynomial},\ }\href {https://doi.org/10.1007/s00453-008-9168-0} {\bibfield  {journal} {\bibinfo  {journal} {Algorithmica}\ }\textbf {\bibinfo {volume} {55}},\ \bibinfo {pages} {395–421} (\bibinfo {year} {2008})}\BibitemShut {NoStop}%
\bibitem [{\citenamefont {Schiansky}\ \emph {et~al.}(2023)\citenamefont {Schiansky}, \citenamefont {Str{\"o}mberg}, \citenamefont {Trillo}, \citenamefont {Saggio}, \citenamefont {Dive}, \citenamefont {Navascu{\'e}s},\ and\ \citenamefont {Walther}}]{schiansky2023demonstration}%
  \BibitemOpen
  \bibfield  {author} {\bibinfo {author} {\bibfnamefont {P.}~\bibnamefont {Schiansky}}, \bibinfo {author} {\bibfnamefont {T.}~\bibnamefont {Str{\"o}mberg}}, \bibinfo {author} {\bibfnamefont {D.}~\bibnamefont {Trillo}}, \bibinfo {author} {\bibfnamefont {V.}~\bibnamefont {Saggio}}, \bibinfo {author} {\bibfnamefont {B.}~\bibnamefont {Dive}}, \bibinfo {author} {\bibfnamefont {M.}~\bibnamefont {Navascu{\'e}s}},\ and\ \bibinfo {author} {\bibfnamefont {P.}~\bibnamefont {Walther}},\ }\bibfield  {title} {\bibinfo {title} {Demonstration of universal time-reversal for qubit processes},\ }\href {https://doi.org/10.1364/OPTICA.469109} {\bibfield  {journal} {\bibinfo  {journal} {Optica}\ }\textbf {\bibinfo {volume} {10}},\ \bibinfo {pages} {200} (\bibinfo {year} {2023})}\BibitemShut {NoStop}%
\bibitem [{\citenamefont {Larocca}\ \emph {et~al.}(2022)\citenamefont {Larocca}, \citenamefont {Sauvage}, \citenamefont {Sbahi}, \citenamefont {Verdon}, \citenamefont {Coles},\ and\ \citenamefont {Cerezo}}]{larocca2022group}%
  \BibitemOpen
  \bibfield  {author} {\bibinfo {author} {\bibfnamefont {M.}~\bibnamefont {Larocca}}, \bibinfo {author} {\bibfnamefont {F.}~\bibnamefont {Sauvage}}, \bibinfo {author} {\bibfnamefont {F.~M.}\ \bibnamefont {Sbahi}}, \bibinfo {author} {\bibfnamefont {G.}~\bibnamefont {Verdon}}, \bibinfo {author} {\bibfnamefont {P.~J.}\ \bibnamefont {Coles}},\ and\ \bibinfo {author} {\bibfnamefont {M.}~\bibnamefont {Cerezo}},\ }\bibfield  {title} {\bibinfo {title} {{Group-Invariant Quantum Machine Learning}},\ }\href {https://doi.org/10.1103/PRXQuantum.3.030341} {\bibfield  {journal} {\bibinfo  {journal} {PRX Quantum}\ }\textbf {\bibinfo {volume} {3}},\ \bibinfo {pages} {030341} (\bibinfo {year} {2022})}\BibitemShut {NoStop}%
\bibitem [{\citenamefont {Fujii}(2003)}]{fujii2003exchange}%
  \BibitemOpen
  \bibfield  {author} {\bibinfo {author} {\bibfnamefont {K.}~\bibnamefont {Fujii}},\ }\bibfield  {title} {\bibinfo {title} {{Exchange gate on the qudit space and Fock space}},\ }\href {https://doi.org/10.1088/1464-4266/5/6/011} {\bibfield  {journal} {\bibinfo  {journal} {J. Opt. B: Quantum Semiclass. Opt.}\ }\textbf {\bibinfo {volume} {5}},\ \bibinfo {pages} {S613} (\bibinfo {year} {2003})}\BibitemShut {NoStop}%
\bibitem [{\citenamefont {Foulds}\ \emph {et~al.}(2024)\citenamefont {Foulds}, \citenamefont {Prove},\ and\ \citenamefont {Kendon}}]{foulds2024generalising}%
  \BibitemOpen
  \bibfield  {author} {\bibinfo {author} {\bibfnamefont {S.}~\bibnamefont {Foulds}}, \bibinfo {author} {\bibfnamefont {O.}~\bibnamefont {Prove}},\ and\ \bibinfo {author} {\bibfnamefont {V.}~\bibnamefont {Kendon}},\ }\bibfield  {title} {\bibinfo {title} {Generalizing multipartite concentratable entanglement for practical applications: mixed, qudit and optical states},\ }\href {https://doi.org/10.1098/rsta.2024.0411} {\bibfield  {journal} {\bibinfo  {journal} {Phil. Trans. R. Soc. A}\ }\textbf {\bibinfo {volume} {382}},\ \bibinfo {pages} {20240411} (\bibinfo {year} {2024})}\BibitemShut {NoStop}%
\bibitem [{\citenamefont {Foulds}\ \emph {et~al.}(2021)\citenamefont {Foulds}, \citenamefont {Kendon},\ and\ \citenamefont {Spiller}}]{foulds2021controlledSWAP}%
  \BibitemOpen
  \bibfield  {author} {\bibinfo {author} {\bibfnamefont {S.}~\bibnamefont {Foulds}}, \bibinfo {author} {\bibfnamefont {V.}~\bibnamefont {Kendon}},\ and\ \bibinfo {author} {\bibfnamefont {T.}~\bibnamefont {Spiller}},\ }\bibfield  {title} {\bibinfo {title} {{The controlled SWAP test for determining quantum entanglement}},\ }\href {https://doi.org/10.1088/2058-9565/abe458} {\bibfield  {journal} {\bibinfo  {journal} {Quantum Sci. Technol.}\ }\textbf {\bibinfo {volume} {6}},\ \bibinfo {pages} {035002} (\bibinfo {year} {2021})}\BibitemShut {NoStop}%
\bibitem [{\citenamefont {Cohen}\ and\ \citenamefont {Pollak}(2018)}]{cohen2018determination}%
  \BibitemOpen
  \bibfield  {author} {\bibinfo {author} {\bibfnamefont {E.}~\bibnamefont {Cohen}}\ and\ \bibinfo {author} {\bibfnamefont {E.}~\bibnamefont {Pollak}},\ }\bibfield  {title} {\bibinfo {title} {Determination of weak values of quantum operators using only strong measurements},\ }\href {https://doi.org/10.1103/PhysRevA.98.042112} {\bibfield  {journal} {\bibinfo  {journal} {Phys. Rev. A}\ }\textbf {\bibinfo {volume} {98}},\ \bibinfo {pages} {042112} (\bibinfo {year} {2018})}\BibitemShut {NoStop}%
\bibitem [{\citenamefont {Nielsen}\ and\ \citenamefont {Chuang}(2010)}]{nielsen2010quantum}%
  \BibitemOpen
  \bibfield  {author} {\bibinfo {author} {\bibfnamefont {M.~A.}\ \bibnamefont {Nielsen}}\ and\ \bibinfo {author} {\bibfnamefont {I.~L.}\ \bibnamefont {Chuang}},\ }\href@noop {} {\emph {\bibinfo {title} {Quantum computation and quantum information}}}\ (\bibinfo  {publisher} {Cambridge university press},\ \bibinfo {year} {2010})\BibitemShut {NoStop}%
\bibitem [{Note2()}]{Note2}%
  \BibitemOpen
  \bibinfo {note} {This is also known as the Fredkin gate.}\BibitemShut {Stop}%
\bibitem [{Note3()}]{Note3}%
  \BibitemOpen
  \bibinfo {note} {Note that for any $\protect \mathbf {j}$ and any $\protect \pmb \varrho $ we have that there exists a value $c \in \protect \mathbb {R}$ and some Bargmann invariant $\Delta _{n}$ such that $\square _{n}(\protect \mathbf {j},\protect \pmb \varrho ) = c \cdot \Delta _{n}$. In other words, $\square _{n}(\protect \mathbf {j},\protect \pmb \varrho )$ is always just a scaling of some Bargmann invariant as given by Eq.~\protect \eqref {eq: Bargmann invariant}. See also the discussion from Ref.~\cite {wagner2024quantum} on the differences between extended Kirkwood--Dirac quasiprobability distributions and Bargmann invariants.}\BibitemShut {Stop}%
\bibitem [{\citenamefont {Reascos}\ \emph {et~al.}(2023)\citenamefont {Reascos}, \citenamefont {Murta}, \citenamefont {Galv\~ao},\ and\ \citenamefont {Fern\'andez-Rossier}}]{reascos2023quantumcircuits}%
  \BibitemOpen
  \bibfield  {author} {\bibinfo {author} {\bibfnamefont {L.~I.}\ \bibnamefont {Reascos}}, \bibinfo {author} {\bibfnamefont {B.}~\bibnamefont {Murta}}, \bibinfo {author} {\bibfnamefont {E.~F.}\ \bibnamefont {Galv\~ao}},\ and\ \bibinfo {author} {\bibfnamefont {J.}~\bibnamefont {Fern\'andez-Rossier}},\ }\bibfield  {title} {\bibinfo {title} {Quantum circuits to measure scalar spin chirality},\ }\href {https://doi.org/10.1103/PhysRevResearch.5.043087} {\bibfield  {journal} {\bibinfo  {journal} {Phys. Rev. Research}\ }\textbf {\bibinfo {volume} {5}},\ \bibinfo {pages} {043087} (\bibinfo {year} {2023})}\BibitemShut {NoStop}%
\bibitem [{\citenamefont {Di\'osi}(2016)}]{diosi2016structural}%
  \BibitemOpen
  \bibfield  {author} {\bibinfo {author} {\bibfnamefont {L.}~\bibnamefont {Di\'osi}},\ }\bibfield  {title} {\bibinfo {title} {Structural features of sequential weak measurements},\ }\href {https://doi.org/10.1103/PhysRevA.94.010103} {\bibfield  {journal} {\bibinfo  {journal} {Phys. Rev. A}\ }\textbf {\bibinfo {volume} {94}},\ \bibinfo {pages} {010103} (\bibinfo {year} {2016})}\BibitemShut {NoStop}%
\bibitem [{\citenamefont {Georgiev}\ and\ \citenamefont {Cohen}(2018)}]{georgiev2018probing}%
  \BibitemOpen
  \bibfield  {author} {\bibinfo {author} {\bibfnamefont {D.}~\bibnamefont {Georgiev}}\ and\ \bibinfo {author} {\bibfnamefont {E.}~\bibnamefont {Cohen}},\ }\bibfield  {title} {\bibinfo {title} {Probing finite coarse-grained virtual {F}eynman histories with sequential weak values},\ }\href {https://doi.org/10.1103/PhysRevA.97.052102} {\bibfield  {journal} {\bibinfo  {journal} {Phys. Rev. A}\ }\textbf {\bibinfo {volume} {97}},\ \bibinfo {pages} {052102} (\bibinfo {year} {2018})}\BibitemShut {NoStop}%
\bibitem [{\citenamefont {Faehrmann}\ \emph {et~al.}(2025)\citenamefont {Faehrmann}, \citenamefont {Eisert},\ and\ \citenamefont {Kueng}}]{faehrmann2025shadowhadamardtestusing}%
  \BibitemOpen
  \bibfield  {author} {\bibinfo {author} {\bibfnamefont {P.~K.}\ \bibnamefont {Faehrmann}}, \bibinfo {author} {\bibfnamefont {J.}~\bibnamefont {Eisert}},\ and\ \bibinfo {author} {\bibfnamefont {R.}~\bibnamefont {Kueng}},\ }\href {https://arxiv.org/abs/2505.15913} {\bibinfo {title} {In the shadow of the {H}adamard test: Using the garbage state for good and further modifications}} (\bibinfo {year} {2025}),\ \Eprint {https://arxiv.org/abs/2505.15913} {arXiv:2505.15913 [quant-ph]} \BibitemShut {NoStop}%
\bibitem [{\citenamefont {Chiribella}\ \emph {et~al.}(2009)\citenamefont {Chiribella}, \citenamefont {D'Ariano}, \citenamefont {Perinotti},\ and\ \citenamefont {Valiron}}]{chiribella2009beyond}%
  \BibitemOpen
  \bibfield  {author} {\bibinfo {author} {\bibfnamefont {G.}~\bibnamefont {Chiribella}}, \bibinfo {author} {\bibfnamefont {G.~M.}\ \bibnamefont {D'Ariano}}, \bibinfo {author} {\bibfnamefont {P.}~\bibnamefont {Perinotti}},\ and\ \bibinfo {author} {\bibfnamefont {B.}~\bibnamefont {Valiron}},\ }\href {https://arxiv.org/abs/0912.0195} {\bibinfo {title} {Beyond quantum computers}} (\bibinfo {year} {2009}),\ \Eprint {https://arxiv.org/abs/0912.0195} {arXiv:0912.0195 [quant-ph]} \BibitemShut {NoStop}%
\bibitem [{\citenamefont {Chiribella}\ \emph {et~al.}(2013)\citenamefont {Chiribella}, \citenamefont {D'Ariano}, \citenamefont {Perinotti},\ and\ \citenamefont {Valiron}}]{Chiribella2013}%
  \BibitemOpen
  \bibfield  {author} {\bibinfo {author} {\bibfnamefont {G.}~\bibnamefont {Chiribella}}, \bibinfo {author} {\bibfnamefont {G.~M.}\ \bibnamefont {D'Ariano}}, \bibinfo {author} {\bibfnamefont {P.}~\bibnamefont {Perinotti}},\ and\ \bibinfo {author} {\bibfnamefont {B.}~\bibnamefont {Valiron}},\ }\bibfield  {title} {\bibinfo {title} {Quantum computations without definite causal structure},\ }\href {https://doi.org/10.1103/PhysRevA.88.022318} {\bibfield  {journal} {\bibinfo  {journal} {Phys. Rev. A}\ }\textbf {\bibinfo {volume} {88}},\ \bibinfo {pages} {022318} (\bibinfo {year} {2013})}\BibitemShut {NoStop}%
\bibitem [{\citenamefont {Chiribella}(2012)}]{Chiribella2012}%
  \BibitemOpen
  \bibfield  {author} {\bibinfo {author} {\bibfnamefont {G.}~\bibnamefont {Chiribella}},\ }\bibfield  {title} {\bibinfo {title} {Perfect discrimination of no-signalling channels via quantum superposition of causal structures},\ }\href {https://doi.org/10.1103/PhysRevA.86.040301} {\bibfield  {journal} {\bibinfo  {journal} {Phys. Rev. A}\ }\textbf {\bibinfo {volume} {86}},\ \bibinfo {pages} {040301(R)} (\bibinfo {year} {2012})}\BibitemShut {NoStop}%
\bibitem [{\citenamefont {Ara\'{u}jo}\ \emph {et~al.}(2014)\citenamefont {Ara\'{u}jo}, \citenamefont {Costa},\ and\ \citenamefont {\v{C}. Brukner}}]{Araujo2014}%
  \BibitemOpen
  \bibfield  {author} {\bibinfo {author} {\bibfnamefont {M.}~\bibnamefont {Ara\'{u}jo}}, \bibinfo {author} {\bibfnamefont {F.}~\bibnamefont {Costa}},\ and\ \bibinfo {author} {\bibnamefont {\v{C}. Brukner}},\ }\bibfield  {title} {\bibinfo {title} {Computational advantage from quantum-controlled ordering of gates},\ }\href {https://doi.org/10.1103/PhysRevLett.113.250402} {\bibfield  {journal} {\bibinfo  {journal} {Phys. Rev. Lett.}\ }\textbf {\bibinfo {volume} {113}},\ \bibinfo {pages} {250402} (\bibinfo {year} {2014})}\BibitemShut {NoStop}%
\bibitem [{\citenamefont {Felce}\ and\ \citenamefont {Vedral}(2020)}]{Felce2020}%
  \BibitemOpen
  \bibfield  {author} {\bibinfo {author} {\bibfnamefont {D.}~\bibnamefont {Felce}}\ and\ \bibinfo {author} {\bibfnamefont {V.}~\bibnamefont {Vedral}},\ }\bibfield  {title} {\bibinfo {title} {Quantum refrigeration with indefinite causal order},\ }\href {https://doi.org/10.1103/PhysRevLett.125.070603} {\bibfield  {journal} {\bibinfo  {journal} {Phys. Rev. Lett.}\ }\textbf {\bibinfo {volume} {125}},\ \bibinfo {pages} {070603} (\bibinfo {year} {2020})}\BibitemShut {NoStop}%
\bibitem [{\citenamefont {Zhao}\ \emph {et~al.}(2020)\citenamefont {Zhao}, \citenamefont {Yang},\ and\ \citenamefont {Chiribella}}]{Zhao2020}%
  \BibitemOpen
  \bibfield  {author} {\bibinfo {author} {\bibfnamefont {X.}~\bibnamefont {Zhao}}, \bibinfo {author} {\bibfnamefont {Y.}~\bibnamefont {Yang}},\ and\ \bibinfo {author} {\bibfnamefont {G.}~\bibnamefont {Chiribella}},\ }\bibfield  {title} {\bibinfo {title} {Quantum metrology with indefinite causal order},\ }\href {https://doi.org/10.1103/PhysRevLett.124.190503} {\bibfield  {journal} {\bibinfo  {journal} {Phys. Rev. Lett.}\ }\textbf {\bibinfo {volume} {124}},\ \bibinfo {pages} {190503} (\bibinfo {year} {2020})}\BibitemShut {NoStop}%
\bibitem [{\citenamefont {Chiribella}\ \emph {et~al.}(2021)\citenamefont {Chiribella}, \citenamefont {Banik}, \citenamefont {Bhattacharya}, \citenamefont {Guha}, \citenamefont {Alimuddin}, \citenamefont {Roy}, \citenamefont {Saha}, \citenamefont {Agrawal},\ and\ \citenamefont {Kar}}]{Chiribella2021}%
  \BibitemOpen
  \bibfield  {author} {\bibinfo {author} {\bibfnamefont {G.}~\bibnamefont {Chiribella}}, \bibinfo {author} {\bibfnamefont {M.}~\bibnamefont {Banik}}, \bibinfo {author} {\bibfnamefont {S.~S.}\ \bibnamefont {Bhattacharya}}, \bibinfo {author} {\bibfnamefont {T.}~\bibnamefont {Guha}}, \bibinfo {author} {\bibfnamefont {M.}~\bibnamefont {Alimuddin}}, \bibinfo {author} {\bibfnamefont {A.}~\bibnamefont {Roy}}, \bibinfo {author} {\bibfnamefont {S.}~\bibnamefont {Saha}}, \bibinfo {author} {\bibfnamefont {S.}~\bibnamefont {Agrawal}},\ and\ \bibinfo {author} {\bibfnamefont {G.}~\bibnamefont {Kar}},\ }\bibfield  {title} {\bibinfo {title} {Indefinite causal order enables perfect quantum communication with zero capacity channels},\ }\href {https://doi.org/10.1088/1367-2630/abe7a0} {\bibfield  {journal} {\bibinfo  {journal} {New J. Phys.}\ }\textbf {\bibinfo {volume} {23}},\ \bibinfo {pages} {033039} (\bibinfo {year} {2021})}\BibitemShut {NoStop}%
\bibitem [{\citenamefont {Wechs}\ \emph {et~al.}(2021)\citenamefont {Wechs}, \citenamefont {Dourdent}, \citenamefont {Abbott},\ and\ \citenamefont {Branciard}}]{Wechs2021}%
  \BibitemOpen
  \bibfield  {author} {\bibinfo {author} {\bibfnamefont {J.}~\bibnamefont {Wechs}}, \bibinfo {author} {\bibfnamefont {H.}~\bibnamefont {Dourdent}}, \bibinfo {author} {\bibfnamefont {A.~A.}\ \bibnamefont {Abbott}},\ and\ \bibinfo {author} {\bibfnamefont {C.}~\bibnamefont {Branciard}},\ }\bibfield  {title} {\bibinfo {title} {Quantum circuits with classical versus quantum control of causal order},\ }\href {https://doi.org/10.1103/PRXQuantum.2.030335} {\bibfield  {journal} {\bibinfo  {journal} {PRX Quantum}\ }\textbf {\bibinfo {volume} {2}},\ \bibinfo {pages} {030335} (\bibinfo {year} {2021})}\BibitemShut {NoStop}%
\bibitem [{\citenamefont {Zhu}\ \emph {et~al.}(2023)\citenamefont {Zhu}, \citenamefont {Chen}, \citenamefont {Hasegawa},\ and\ \citenamefont {Xue}}]{Zhu2021}%
  \BibitemOpen
  \bibfield  {author} {\bibinfo {author} {\bibfnamefont {G.}~\bibnamefont {Zhu}}, \bibinfo {author} {\bibfnamefont {Y.}~\bibnamefont {Chen}}, \bibinfo {author} {\bibfnamefont {Y.}~\bibnamefont {Hasegawa}},\ and\ \bibinfo {author} {\bibfnamefont {P.}~\bibnamefont {Xue}},\ }\bibfield  {title} {\bibinfo {title} {Charging quantum batteries via indefinite causal order: Theory and experiment},\ }\href {https://doi.org/10.1103/PhysRevLett.131.240401} {\bibfield  {journal} {\bibinfo  {journal} {Phys. Rev. Lett.}\ }\textbf {\bibinfo {volume} {131}},\ \bibinfo {pages} {240401} (\bibinfo {year} {2023})}\BibitemShut {NoStop}%
\bibitem [{\citenamefont {Bavaresco}\ \emph {et~al.}(2021)\citenamefont {Bavaresco}, \citenamefont {Murao},\ and\ \citenamefont {Quintino}}]{Bavaresco2021}%
  \BibitemOpen
  \bibfield  {author} {\bibinfo {author} {\bibfnamefont {J.}~\bibnamefont {Bavaresco}}, \bibinfo {author} {\bibfnamefont {M.}~\bibnamefont {Murao}},\ and\ \bibinfo {author} {\bibfnamefont {M.~T.}\ \bibnamefont {Quintino}},\ }\bibfield  {title} {\bibinfo {title} {Strict hierarchy between parallel, sequential, and indefinite-causal-order strategies for channel discrimination},\ }\href {https://doi.org/10.1103/PhysRevLett.127.200504} {\bibfield  {journal} {\bibinfo  {journal} {Phys. Rev. Lett.}\ }\textbf {\bibinfo {volume} {127}},\ \bibinfo {pages} {200504} (\bibinfo {year} {2021})}\BibitemShut {NoStop}%
\bibitem [{\citenamefont {Simonov}\ \emph {et~al.}(2022{\natexlab{a}})\citenamefont {Simonov}, \citenamefont {Francica}, \citenamefont {Guarnieri},\ and\ \citenamefont {Paternostro}}]{Simonov2022}%
  \BibitemOpen
  \bibfield  {author} {\bibinfo {author} {\bibfnamefont {K.}~\bibnamefont {Simonov}}, \bibinfo {author} {\bibfnamefont {G.}~\bibnamefont {Francica}}, \bibinfo {author} {\bibfnamefont {G.}~\bibnamefont {Guarnieri}},\ and\ \bibinfo {author} {\bibfnamefont {M.}~\bibnamefont {Paternostro}},\ }\bibfield  {title} {\bibinfo {title} {Work extraction from coherently activated maps via quantum switch},\ }\href {https://doi.org/10.1103/PhysRevA.105.032217} {\bibfield  {journal} {\bibinfo  {journal} {Phys. Rev. A}\ }\textbf {\bibinfo {volume} {105}},\ \bibinfo {pages} {032217} (\bibinfo {year} {2022}{\natexlab{a}})}\BibitemShut {NoStop}%
\bibitem [{\citenamefont {Simonov}\ \emph {et~al.}(2022{\natexlab{b}})\citenamefont {Simonov}, \citenamefont {Roy}, \citenamefont {Guha}, \citenamefont {Zimborás},\ and\ \citenamefont {Chiribella}}]{Simonov2022_Erg}%
  \BibitemOpen
  \bibfield  {author} {\bibinfo {author} {\bibfnamefont {K.}~\bibnamefont {Simonov}}, \bibinfo {author} {\bibfnamefont {S.}~\bibnamefont {Roy}}, \bibinfo {author} {\bibfnamefont {T.}~\bibnamefont {Guha}}, \bibinfo {author} {\bibfnamefont {Z.}~\bibnamefont {Zimborás}},\ and\ \bibinfo {author} {\bibfnamefont {G.}~\bibnamefont {Chiribella}},\ }\href {https://arxiv.org/abs/2208.04034} {\bibinfo {title} {Activation of thermal states by coherently controlled thermalization processes}} (\bibinfo {year} {2022}{\natexlab{b}}),\ \Eprint {https://arxiv.org/abs/2208.04034} {arXiv:2208.04034 [quant-ph]} \BibitemShut {NoStop}%
\bibitem [{\citenamefont {Koudia}\ \emph {et~al.}(2022)\citenamefont {Koudia}, \citenamefont {Cacciapuoti}, \citenamefont {Simonov},\ and\ \citenamefont {Caleffi}}]{Koudia2022}%
  \BibitemOpen
  \bibfield  {author} {\bibinfo {author} {\bibfnamefont {S.}~\bibnamefont {Koudia}}, \bibinfo {author} {\bibfnamefont {A.~S.}\ \bibnamefont {Cacciapuoti}}, \bibinfo {author} {\bibfnamefont {K.}~\bibnamefont {Simonov}},\ and\ \bibinfo {author} {\bibfnamefont {M.}~\bibnamefont {Caleffi}},\ }\bibfield  {title} {\bibinfo {title} {How deep the theory of quantum communications goes: Superadditivity, superactivation and causal activation},\ }\href {https://doi.org/10.1109/COMST.2022.3196449} {\bibfield  {journal} {\bibinfo  {journal} {IEEE Comm. Surv. Tutor.}\ }\textbf {\bibinfo {volume} {24}},\ \bibinfo {pages} {1926} (\bibinfo {year} {2022})}\BibitemShut {NoStop}%
\bibitem [{\citenamefont {Liu}\ \emph {et~al.}(2023)\citenamefont {Liu}, \citenamefont {Hu}, \citenamefont {Yuan},\ and\ \citenamefont {Yang}}]{Liu2023}%
  \BibitemOpen
  \bibfield  {author} {\bibinfo {author} {\bibfnamefont {Q.}~\bibnamefont {Liu}}, \bibinfo {author} {\bibfnamefont {Z.}~\bibnamefont {Hu}}, \bibinfo {author} {\bibfnamefont {H.}~\bibnamefont {Yuan}},\ and\ \bibinfo {author} {\bibfnamefont {Y.}~\bibnamefont {Yang}},\ }\bibfield  {title} {\bibinfo {title} {Optimal strategies of quantum metrology with a strict hierarchy},\ }\href {https://doi.org/10.1103/PhysRevLett.130.070803} {\bibfield  {journal} {\bibinfo  {journal} {Phys. Rev. Lett.}\ }\textbf {\bibinfo {volume} {130}},\ \bibinfo {pages} {070803} (\bibinfo {year} {2023})}\BibitemShut {NoStop}%
\bibitem [{\citenamefont {Caleffi}\ \emph {et~al.}(2023)\citenamefont {Caleffi}, \citenamefont {Simonov},\ and\ \citenamefont {Cacciapuoti}}]{CalSimCac-23}%
  \BibitemOpen
  \bibfield  {author} {\bibinfo {author} {\bibfnamefont {M.}~\bibnamefont {Caleffi}}, \bibinfo {author} {\bibfnamefont {K.}~\bibnamefont {Simonov}},\ and\ \bibinfo {author} {\bibfnamefont {A.~S.}\ \bibnamefont {Cacciapuoti}},\ }\bibfield  {title} {\bibinfo {title} {Beyond {S}hannon limits: Quantum communications through quantum paths},\ }\href {https://doi.org/10.1109/JSAC.2023.3288263} {\bibfield  {journal} {\bibinfo  {journal} {IEEE J. Sel. Areas Commun.}\ }\textbf {\bibinfo {volume} {41}},\ \bibinfo {pages} {2707} (\bibinfo {year} {2023})}\BibitemShut {NoStop}%
\bibitem [{\citenamefont {Spencer-Wood}(2023)}]{HectorWood2023}%
  \BibitemOpen
  \bibfield  {author} {\bibinfo {author} {\bibfnamefont {H.}~\bibnamefont {Spencer-Wood}},\ }\href {https://arxiv.org/abs/2303.03893} {\bibinfo {title} {Indefinite causal key distribution}} (\bibinfo {year} {2023}),\ \Eprint {https://arxiv.org/abs/2303.03893} {arXiv:2303.03893 [quant-ph]} \BibitemShut {NoStop}%
\bibitem [{\citenamefont {Simonov}\ \emph {et~al.}(2023)\citenamefont {Simonov}, \citenamefont {Caleffi}, \citenamefont {Illiano}, \citenamefont {Romero},\ and\ \citenamefont {Cacciapuoti}}]{Simonov2023}%
  \BibitemOpen
  \bibfield  {author} {\bibinfo {author} {\bibfnamefont {K.}~\bibnamefont {Simonov}}, \bibinfo {author} {\bibfnamefont {M.}~\bibnamefont {Caleffi}}, \bibinfo {author} {\bibfnamefont {J.}~\bibnamefont {Illiano}}, \bibinfo {author} {\bibfnamefont {J.}~\bibnamefont {Romero}},\ and\ \bibinfo {author} {\bibfnamefont {A.~S.}\ \bibnamefont {Cacciapuoti}},\ }\href {https://arxiv.org/abs/2311.13654} {\bibinfo {title} {Universal quantum computation via superposed orders of single-qubit gates}} (\bibinfo {year} {2023}),\ \Eprint {https://arxiv.org/abs/2311.13654} {arXiv:2311.13654 [quant-ph]} \BibitemShut {NoStop}%
\bibitem [{\citenamefont {Liu}\ \emph {et~al.}(2024)\citenamefont {Liu}, \citenamefont {Meng}, \citenamefont {Song}, \citenamefont {Li}, \citenamefont {Wu}, \citenamefont {Chen}, \citenamefont {Hong}, \citenamefont {Zhang},\ and\ \citenamefont {Yin}}]{Liu2024_Deutsch}%
  \BibitemOpen
  \bibfield  {author} {\bibinfo {author} {\bibfnamefont {W.-Q.}\ \bibnamefont {Liu}}, \bibinfo {author} {\bibfnamefont {Z.}~\bibnamefont {Meng}}, \bibinfo {author} {\bibfnamefont {B.-W.}\ \bibnamefont {Song}}, \bibinfo {author} {\bibfnamefont {J.}~\bibnamefont {Li}}, \bibinfo {author} {\bibfnamefont {Q.-Y.}\ \bibnamefont {Wu}}, \bibinfo {author} {\bibfnamefont {X.-X.}\ \bibnamefont {Chen}}, \bibinfo {author} {\bibfnamefont {J.-Y.}\ \bibnamefont {Hong}}, \bibinfo {author} {\bibfnamefont {A.-N.}\ \bibnamefont {Zhang}},\ and\ \bibinfo {author} {\bibfnamefont {Z.-Q.}\ \bibnamefont {Yin}},\ }\bibfield  {title} {\bibinfo {title} {Experimentally demonstrating indefinite causal order algorithms to solve the generalized {D}eutsch's problem},\ }\href {https://doi.org/https://doi.org/10.1002/qute.202400181} {\bibfield  {journal} {\bibinfo  {journal} {Adv. Quantum Technol.}\ }\textbf {\bibinfo {volume} {7}},\ \bibinfo {pages} {2400181} (\bibinfo {year} {2024})}\BibitemShut {NoStop}%
\bibitem [{\citenamefont {Rozema}\ \emph {et~al.}(2024)\citenamefont {Rozema}, \citenamefont {Str\"omberg}, \citenamefont {Cao}, \citenamefont {Guo}, \citenamefont {Liu},\ and\ \citenamefont {Walther}}]{Rozema2024}%
  \BibitemOpen
  \bibfield  {author} {\bibinfo {author} {\bibfnamefont {L.}~\bibnamefont {Rozema}}, \bibinfo {author} {\bibfnamefont {T.}~\bibnamefont {Str\"omberg}}, \bibinfo {author} {\bibfnamefont {H.}~\bibnamefont {Cao}}, \bibinfo {author} {\bibfnamefont {Y.}~\bibnamefont {Guo}}, \bibinfo {author} {\bibfnamefont {B.-H.}\ \bibnamefont {Liu}},\ and\ \bibinfo {author} {\bibfnamefont {P.}~\bibnamefont {Walther}},\ }\bibfield  {title} {\bibinfo {title} {Experimental aspects of indefinite causal order in quantum mechanics},\ }\href {https://doi.org/10.1038/s42254-024-00739-8} {\bibfield  {journal} {\bibinfo  {journal} {Nat. Rev. Phys.}\ }\textbf {\bibinfo {volume} {6}},\ \bibinfo {pages} {483} (\bibinfo {year} {2024})}\BibitemShut {NoStop}%
\bibitem [{\citenamefont {Swingle}\ \emph {et~al.}(2016)\citenamefont {Swingle}, \citenamefont {Bentsen}, \citenamefont {Schleier-Smith},\ and\ \citenamefont {Hayden}}]{Swingle2016}%
  \BibitemOpen
  \bibfield  {author} {\bibinfo {author} {\bibfnamefont {B.}~\bibnamefont {Swingle}}, \bibinfo {author} {\bibfnamefont {G.}~\bibnamefont {Bentsen}}, \bibinfo {author} {\bibfnamefont {M.}~\bibnamefont {Schleier-Smith}},\ and\ \bibinfo {author} {\bibfnamefont {P.}~\bibnamefont {Hayden}},\ }\bibfield  {title} {\bibinfo {title} {Measuring the scrambling of quantum information},\ }\href {https://doi.org/10.1103/PhysRevA.94.040302} {\bibfield  {journal} {\bibinfo  {journal} {Phys. Rev. A}\ }\textbf {\bibinfo {volume} {94}},\ \bibinfo {pages} {040302(R)} (\bibinfo {year} {2016})}\BibitemShut {NoStop}%
\end{thebibliography}%

\end{document}